\documentclass{LMCS}

\usepackage{xy}
\xyoption{all}
\usepackage{prooftree}
\usepackage{amssymb}

\usepackage{hyperref}

\newcommand{\key}[1]{{\sf #1}}
\renewcommand{\ifthenelse}[3]{\key{if}\; #1 \;\key{then}\; #2 \;\key{else}\; #3}
\newcommand{\Ifthenelse}[3]{\begin{array}[t]{@{}l}
                \key{if}\; #1\;\, \\
                \key{then}\; #2 \\
                \key{else}\; #3
                \end{array}}

\newcommand{\type}{{\sf Type}}
\newcommand{\CPO}{{\sf CPO}}

\newcommand{\comm}{{\sf comm}}

\newcommand{\Set}{{\sf Set}}
\newcommand{\FIV}{{\sf fv}}

\newcommand{\emp}{{\sf emp}}
\newcommand{\true}{{\sf true}}
\newcommand{\false}{{\sf false}}
\def\correct#1{\relax \ifmmode \expandafter #1 \else $\expandafter #1$\fi}            
\def\mimp{\mbox{$\correct {-\!\!*}\,$}}

\def\doframeit#1{\vbox{%
  \hrule height\fboxrule
    \hbox{%
      \vrule width\fboxrule \kern\fboxsep
      \vbox{\kern\fboxsep #1\kern\fboxsep }%
      \kern\fboxsep \vrule width\fboxrule }%
    \hrule height\fboxrule }}

\def\frameit{\smallskip \advance \linewidth by -7.5pt \setbox0=\vbox \bgroup
\strut \ignorespaces }

\def\endframeit{\ifhmode \par \nointerlineskip \fi \egroup
\doframeit{\box0}}

\newcommand{\new}{{\sf new}}
\newcommand{\free}{{\sf free}}

\newcommand{\mylet}[2]{{{\sf let}\,{#1}\,{\sf in}\,{#2}}}

\newcommand{\myskip}{{\sf skip}}
\newcommand{\ifz}{{\sf ifz}}
\newcommand{\fix}{{\sf fix}}
\newcommand{\lfix}{{\sf lfix}}

\newcommand{\IVar}{{\sf Ids}}

\newcommand{\trity}[2]{{\{{#1}\}\mbox{-}\{{#2}\}}}
\newcommand{\bind}{{\scriptstyle\relax\rightarrow}}

\newcommand{\T}{{\bf T}}
\newcommand{\CC}{{\mathcal{C}}}
\newcommand{\cD}{{\mathcal{D}}}
\newcommand{\cP}{{\mathcal{P}}}
\newcommand{\cPr}{{\mathcal{P}}_r}
\newcommand{\inv}{{\sf inv}}
\newcommand{\tri}{{\sf tri}}

\newcommand{\ff}[1]{[\![#1]\!]}
\newcommand{\cff}[1]{[\![#1]\!]^\CC}
\newcommand{\sff}[1]{[\![#1]\!]^\CPO}

\newcommand{\defeq}{\stackrel{\mbox{\rm\scriptsize def}}{=}}

\newcommand{\Id}{{\sf Id}}

\newcommand{\sskip}{\mathit{skip}}
\newcommand{\sseq}{\mathit{seq}}
\newcommand{\sfree}{\mathit{free}}
\newcommand{\swrite}{\mathit{write}}
\newcommand{\snew}{\mathit{new}}
\newcommand{\sread}{\mathit{read}}

\newcommand{\sloc}{\mathit{Loc}}
\newcommand{\sval}{\mathit{Val}}

\newcommand{\undefined}{\mathit{undef}}
\newcommand{\strue}{\mathit{true}}
\newcommand{\semp}{\mathit{emp}}
\newcommand{\sint}{\mathit{Int}}
\newcommand{\snat}{\mathit{PositiveInt}}
\newcommand{\spred}{\mathit{Pred}}
\newcommand{\heap}{\mathit{Heap}}
\newcommand{\state}{\mathit{State}}
\newcommand{\Fam}{\mathit{Fam}}
\newcommand{\wrong}{\mathit{wrong}}

\newcommand{\curry}{{\sf curry}}
\newcommand{\ev}{{\sf ev}}
\newcommand{\id}{{\sf id}}

\newcommand{\op}{{\sf op}}

\newcommand{\dom}{{\sf dom}}
\newcommand{\fin}{{\sf fin}}

\newcommand{\wwedge}{\;{\wedge}\;}

\newcommand{\botcomp}{\mathit{CR}}

\newcommand{\myfst}{{\sf fst}}

\newcommand{\commentout}[1]{}

\newcommand{\lst}{{\sf lst}}

\newcommand{\dlist}{\mathit{Dlist}}

\newcommand{\Malloc}{\mathit{Malloc}}
\newcommand{\Mfree}{\mathit{Mfree}}
\newcommand{\Body}{\mathit{Body}}

\newcommand{\FalseBranch}{\mathit{FBranch}}
\newcommand{\Rand}{\mathit{Rd}}
\newcommand{\Client}{\mathit{Client}}

\newcommand{\coinflip}{\mathit{cflip}}

\newcommand{\mfree}{\mathit{mfree}}

\newcommand{\con}{\mathit{con}}

\def\doi{2 (5:1) 2006}
\lmcsheading%
{\doi}
{1--33}
{}
{}
{Dec.~21, 2005}
{Nov.~\phantom{0}3, 2006}
{}   

\begin{document}

\title[Semantics of Separation-logic Typing and Higher-order Frame Rules]
   {Semantics of Separation-logic Typing and Higher-order Frame Rules 
    for Algol-like Languages\rsuper *}

\author[L.~Birkedal]{Lars Birkedal\rsuper a}
\address{{\lsuper{a,b}}IT University of Copenhagen, Denmark}
\email{\{birkedal,noah\}@itu.dk}

\author[N.~Torp-Smith]{Noah Torp-Smith\rsuper b}
\address{\vskip-8 pt}

\author[H.~Yang]{Hongseok Yang\rsuper c}
\address{{\lsuper c}Seoul National University, Korea}
\email{hyang@ropas.snu.ac.kr}

\subjclass{F.3, D.3}
\keywords{Program Verification, Separation Logic}
\titlecomment{{\lsuper *}An extended abstract of this paper appeared at LICS'05}
\begin{abstract}
     We show how to give a coherent semantics to programs that are
     well-specified in a version of separation logic for a language
     with higher types: idealized algol extended with heaps (but
     with immutable stack variables). 
     In particular, we provide simple sound rules for deriving
     higher-order frame rules, allowing for local reasoning.
\end{abstract}

\maketitle

\section{Introduction}
\label{sec:intro}

Separation
logic~\cite{reynolds02,reynolds00,
ohearn-ishtiaq-popl01,yang-reddy04,ohearn-reynolds-yang01,birkedal-torpsmith-reynolds-popl04}
is a Hoare-style program logic, and variants of it have been applied
to prove correct interesting pointer algorithms such as copying a dag,
disposing a graph, the Schorr-Waite graph algorithm, and Cheney's
copying garbage collector.  The main advantage of separation logic
compared to ordinary Hoare logic is that it facilitates \emph{local
  reasoning}, formalized via the so-called \emph{frame rule} using a
connective called \emph{separating conjunction}.  The development of
separation logic has mostly focused on \emph{low-level} languages with
heaps and pointers, although in recent
work~\cite{yang-ohearn-reynolds-popl04} it was shown how to extend
separation logic to a language with a simple kind of procedures, and a
second-order frame rule was proved sound.

Our aim here is to extend the study of separation logic to
\emph{high-level} languages, in particular to \emph{higher-order}
languages, in such a way that a wide collection of frame rules are
sound, thus allowing for local reasoning in the presence of
higher-order procedures.  For concreteness, we choose to focus on the
language of idealized algol extended with heaps and pointers and we
develop a semantics for this language in which all commands and
procedures are appropriately local. Our approach is to refine the type
system of idealized algol extended with heaps, essentially by making
specifications be types, and give semantics to well-specified
programs.  Thus we develop a \emph{separation-logic type system} for
idealized algol extended with heaps. It is a dependent type theory and
the types include Hoare triples, rules corresponding to the rules of
separation logic, and subtyping rules formalizing higher-order
versions of the frame rule of separation logic.

Our type system is related to modern proposals for type systems for
low-level imperative languages, such as
TAL~\cite{morriset-walker-crary-glew99:f_to_tal}, in that types may
express state changes (since they include forms of Hoare triples as
types).  The type system for TAL was proved sound using an operational
semantics.  We provide a soundness proof of our type system using a
denotational semantics which we, moreover, formally relate to the
standard semantics for idealized
algol~\cite{oles-thesis,reynolds-essence}.  The denotational semantics
of a well-typed program is given by induction on its typing derivation
and the relation to the standard semantics for idealized algol is then
used to prove that the semantics is \emph{coherent} (i.e., is
independent of the chosen typing derivation).  We should perhaps
stress that soundness is not a trivial issue: Reynolds has
shown~\cite{yang-ohearn-reynolds-popl04} that already the soundness of
the second-order frame rule is tricky, by proving that if a proof
system contains the second-order frame rule and the conjunction rule,
together with the ordinary frame rule and the rule of Consequence, then
the system becomes inconsistent. The semantics of our system proves
that if we drop the conjunction rule, then we get soundness of all
higher-order frame rules, including the second-order one. We also
show how to get soundness of all higher-order frame rules \emph{without}
dropping the conjunction rule, by instead restricting attention to
so-called \emph{precise} predicates (see Section~\ref{sec:conjunction-rule}).

In idealized algol, variables are allocated on a stack and they are
mutable (i.e., one can assign to variables). We only consider
\emph{immutable} variables (as in the ML programming language) for
simplicity.  The reason for this choice is that all mutation then
takes place in the heap and thus we need not bother with so-called
{\em modifies clauses\/} on frame rules, which become complicated to
state already for the second-order frame
rule~\cite{yang-ohearn-reynolds-popl04}.

We now give an intuitive overview of the technical development.
Recall that the standard semantics of idealized algol is given using
the category $\CPO$ of pointed complete partial orders and continuous
functions.  Thus types are interpreted as pointed complete partial
orders and terms (programs) are interpreted as continuous functions.
The semantics of our refined type system is given by refining the
standard semantics.  A type $\theta$ in our refined type system
specifies which elements of the ``underlying'' type in the standard
semantics satisfy the specification corresponding to $\theta$ and are
appropriately local (to ensure soundness of the frame rules), that is,
it ``extracts'' those elements.  Moreover, the semantics also
equates elements, which cannot be distinguished by clients, that is,
it quotients some of the extracted elements. Corresponding to these
two aspects of the semantics we introduce two categories, $\CC$ and
$\cD$, where $\CC$ just contains the extracted elements and $\cD$
is a quotient of $\CC$. Thus there is a faithful functor from $\CC$ to
$\CPO$ and a full functor from $\CC$ to $\cD$. We show that the
categories $\CC$ and $\cD$ are cartesian closed and have additional
structure to interpret the higher-order frame rules, and that the
mentioned functors preserve all this structure. The semantics of our
type system is then given in the category $\cD$ and the functors
relating $\CC$, $\cD$, and $\CPO$ are then used to prove coherence of
the semantics.  In fact, as mentioned above, our type system is a
\emph{dependent} type theory, with dependent product type $\Pi_{i}
\theta$ intuitively corresponding to the specification given by
universally quantifying $i$ in the specification corresponding to
$\theta$ (the usual Curry-Howard correspondence). For this reason the
semantics is really not given in $\cD$ but rather in the family
fibration $\Fam(\cD)\rightarrow\Set$ over $\cD$.

The remainder of this paper is organized as follows. In Section
\ref{sec:storage-model}, we define the storage model and assertion
language used in this paper, thus setting the stage for our model. In
Section \ref{sec:programming-language}, we provide the syntax of the
version of idealized algol we use in this paper. In particular, we
introduce our separation-logic type system, which includes an extended
subtype relation. We also include two extended examples of typings in 
our typing system, one of which exemplifies the use of 
a third-order frame rule.
In Section \ref{sec:semantics}, we present the
main contribution of the paper, a model which allows a sound
interpretation, which we also show to be coherent and in harmony with
the standard semantics.  For simplicity, we omit treatment of the
conjunction rule in Sections~\ref{sec:programming-language}
and~\ref{sec:semantics} --- in Section~\ref{sec:conjunction-rule} we show
how to treat the conjunction rule. In the last sections we give pointers to
related and future work, and conclude.

An extended abstract of this paper was presented at the LICS 2005
conference. Compared to the conference paper, the present paper includes 
proofs, more detailed examples of the use of the typing system, 
and a treatment of the conjunction rule.

\section{Storage Model and Assertion Language}
\label{sec:storage-model}
We use the usual storage model of separation logic with one minor
modification: we make explicit the shape of stack storage.  Let $\IVar
= \{i,j,\ldots\}$ be a countably infinite set of variables, and let
$\Delta$ range over finite subsets of $\IVar$. We use the following
semantic domains:
$$
\begin{array}{rcrcl}
& & \sloc & \defeq &  \snat,  \qquad\qquad\quad
\\
& & \sval & \defeq &  \sint, \qquad\qquad\quad
\\
\eta & \in & \ff{\Delta} & \defeq & \Delta \rightarrow \sval, \\
   h & \in &  \heap & \defeq & \sloc \rightharpoonup_\fin \sval, \\
(\eta,h) & \in &  \state(\Delta) & \defeq & \ff{\Delta} \times \heap.
\end{array}
$$
In this storage model, locations are positive integers, so that
they can be manipulated by arithmetic operations.
The set $\Delta$ models the set of variables in scope, and an element
$\eta$ in $\ff{\Delta}$ specifies the values of those stack variables. We
sometimes call $\eta$ an {\em environment\/} instead of a {\em stack}, in
order to emphasize that all variables are immutable.  An element $h$ in
$\heap$ denotes a heap; the domain of $h$ specifies the set of allocated
cells, and the actual action of $h$ determines the contents of those
allocated cells.  We recall the disjointness predicate $h\#h'$ and the
(partial) heap combination operator $h\cdot h'$ from separation logic.
The predicate $h\#h'$ means that $\dom(h) \cap \dom(h') = \emptyset$; and,
$h\cdot h'$ is defined only for such disjoint heaps $h$ and $h'$, and in
that case, it denotes the combined heap $h\cup h'$.

Properties of states are expressed using the assertion language 
of classical separation logic \cite{reynolds02}:
\footnote{The assertion language of separation logic also
contains the separating implication $\mimp$. Since that connective
does not raise any new issues in connection with the present work,
we omit it here.}
$$
\begin{array}{rcl}
    E  & ::= & i \mid 0 \mid 1 \mid E+E \mid E-E, \\ 
    P  & ::= & E=E 
               \mid E\mapsto E
               \mid \emp
               \mid P * P 
               \mid \true 
               \mid P \wedge P 
               \mid P \vee P 
               \mid \neg P
               \mid \forall i.\, P.
               \mid \exists i.\, P.
\end{array}
$$
The assertion $E\mapsto E'$ means that the current heap has only one
cell $E$ and, moreover, that the content of the cell is $E'$. When
we do not care about the contents, we write $E\mapsto -$;
formally, this is an abbreviation of $\exists i.\, E\mapsto i$
for some $i$ not occurring in $E$.  The next two assertions,
$\emp$ and $P*Q$, are the most interesting features of this assertion language.
The empty predicate $\emp$ means that the current heap is empty, and the
separating conjunction $P*Q$ means that the current heap can be 
partitioned into two parts, one satisfying $P$ and another 
satisfying $Q$.

As in the storage model, we make explicit which set
of free variables we are considering an expression or an 
assertion under.
Thus, letting $\FIV$ be a function that takes
an expression or an assertion and returns the set of
free variables, we often write assertions
as $\Delta \vdash P$ to indicate that 
$\FIV(P) \subseteq \Delta$, and that $P$ is currently being 
considered for environments of the shape $\Delta$. Likewise, we
often write $\Delta \vdash E$ for expressions.

The interpretations of an expression $\Delta \vdash E$ and 
an assertion $\Delta \vdash P$ are of the forms
$$
\ff{\Delta \vdash E} : \ff{\Delta} \rightarrow \sval,
\qquad
\ff{\Delta \vdash P} : \ff{\Delta} \rightarrow \cP(\heap).
$$
The interpretation of expressions is standard, just like that of
assertions. We include part of the definition of the interpretation of
assertions here.
\begin{displaymath}
  \begin{array}{rcl}
    \ff{\Delta \vdash E \mapsto E'}_\eta &=& 
    \ifthenelse
       {(\ff{\Delta \vdash E}_\eta \leq 0)}
       {\emptyset}
       {\bigl\{[\ff{\Delta \vdash E}_\eta \bind \ff{\Delta \vdash E'}_\eta ]\bigr\}},
    \\
    \ff{\Delta \vdash \emp }_\eta  &=&  \{{[]}\},
    \\
    \ff{\Delta \vdash P * P'}_\eta & = &
    \{h\cdot h' \mid h\#h' \land h \in \ff{\Delta \vdash P}_\eta 
    \land h' \in \ff{\Delta \vdash P'}_\eta \},
    \\
    \ff{\Delta \vdash \forall i.\ P}_\eta & = &
     \{h \mid \forall n\in
     \sval.\ h \in \ff{\Delta \cup \{i\} \vdash  P}_{\eta[i\bind n]}\}. 
  \end{array}  
\end{displaymath}


\section{Programming Language}
\label{sec:programming-language}
The programming language is Reynolds's idealized algol
\cite{reynolds-essence} adapted for ``separation-logic typing.'' It is
a call-by-name typed lambda calculus, extended with heap operations,
dependent functions, and Hoare-triple types.  As explained in the
introduction, we only consider immutable variables.


The types of the language are defined as follows. We write
$\Delta\vdash \theta : \type$ for a type $\theta$ in context $\Delta$.
The set of types is defined by the following inference rules
(in which $P$ and $Q$ range over assertions):
$$ 
\begin{array}{cc}
\prooftree
\Delta \vdash P
\quad
\Delta \vdash Q
\justifies
\Delta\vdash\trity{P}{Q} :\type
\endprooftree
&
\prooftree
\Delta\vdash\theta : \type 
\quad 
\Delta \vdash P
\justifies
\Delta\vdash\theta\otimes P : \type
\endprooftree
\\\\
\prooftree
\Delta\cup\{i\}\vdash\theta : \type
\using
(i\notin\Delta)
\justifies
\Delta\vdash\Pi_{i}\theta : \type
\endprooftree
&
\prooftree
\Delta\vdash\theta : \type\quad \Delta\vdash\theta' : \type
\justifies
\Delta\vdash\theta\rightarrow\theta' : \type
\endprooftree
\end{array}
$$


Note that the types are \emph{dependent types}, in that they may depend on
variables $i$ (see the first rule above).
One way to understand a type is to read it as a specification for terms,
i.e., through the Curry-Howard correspondence.  A Hoare-triple type
$\trity{P}{Q}$ is a direct import from separation logic; it denotes a set
of commands $c$ that satisfy the Hoare triple $\{P\}c\{Q\}$.  An invariant
extension $\theta\otimes P$ is satisfied by a term $M$ if and only if 
for one part of the heap, the behavior of $M$ satisfies $\theta$ and 
for the other part of the heap, $M$ maintains the invariant $P$. 
We remark that $\theta\otimes P$ allows $M$ to transfer
cells between the $\theta$-part of the heap
and the $P$-part of the heap. For instance,
$\trity{P}{Q} \otimes P_0$ intuitively consists of the following
commands $c$: given an input state satisfying $P*P_0$, so that the 
input state may be split into a $P$-part and a $P_0$-part, 
command $c$ changes these two parts, sometimes transferring 
cells between the two, such that in
the end, the $P$-part satisfies $Q$ and 
the $P_0$-part satisfies $P_0$.

The type $\Pi_i\theta$ is a dependent product type, as in standard
dependent type theory (under Curry-Howard it corresponds to 
the specification given by universally quantifying
$i$ in the specification corresponding to $\theta$).
Intuitively, $\Pi_{i}\theta$ denotes functions from integers such that
given an integer $n$, they return a value satisfying $\theta[n/i]$. For
example, the type $\Pi_i \trity{j\mapsto -}{j \mapsto i!}$ specifies a
factorial function that computes the factorial of $i$ and
stores the result in the heap cell $j$.
 
The pre-terms of the language are given by the following grammar:
$$
\begin{array}{@{}r@{\,}c@{\,}l@{}}
 M    & ::=  & x \mid \lambda x\colon\theta.M \mid M M 
               \mid \lambda i.M \mid M E \\
      & \mid & \fix\, M \mid \ifz\, E\, M\, M \mid \myskip \mid M;M \\
      & \mid & \mylet{i=\new}{M} \mid
               \free(E) \mid [E]:=E \mid \mylet{i=[E]}{M}, \\
\end{array} 
$$
where $E$ is an integer expression defined in
Section~\ref{sec:storage-model}. The language has the usual constructs
for a higher-order imperative language with heap operations, but it
has two distinct features. First, it treats the integer expressions as
``second class'': the terms $M$ never have the integer type, and all
integer expressions inside a term are from the separate grammar for
$E$ defined in Section~\ref{sec:storage-model}.  Second, no ``integer
variables'' $i$ can be modified in this language; only heap cells can
be modified.  Note that the language has two forms of abstraction and
application, one for general terms and the other for integer
expressions. A consequence of this stratification is that all
integer expressions terminate, because the grammar for $E$ does not
contain the recursion operator.

The language has four heap operations. Command $\mylet{i=\new}{M}$
allocates a heap cell, binds $i$ to the address of the allocated cell, and
executes the command $M$.\footnote{We consider single-cell allocation only
  in order to simplify the presentation; it is straightforward to adapt our
  results to a language with allocation of $n$ consecutive cells.} An
allocated cell $i$ can be disposed by $\free(i)$. The remaining two
commands access the content of a cell. The command $[i]:=E'$ changes the
content of cell $i$ by $E'$; and $\mylet{j=[i]}{M}$ reads the content of
cell $i$, binds $j$ to the read value, and executes $M$. Note that the
allocation and lookup commands involve the ``continuation'', and make the
bound variable available in the continuation; such indirect-style
commands are needed because all variables are immutable.

In this paper, we assume a hygiene condition on integer variables 
$i$, in order to avoid the (well-known) issue of
variable capturing. That is, we assume that no terms
or types in the paper use a single symbol $i$ 
for more than one bound variables, or for a bound variable and a free
variable at the same time. 

The typing rules of the language decide a judgment of the form 
$\Gamma \vdash_\Delta M : \theta$, where $\Gamma$ is a list
of type assignments to identifiers $\Gamma=x_1\colon\theta_1,\ldots,x_n\colon\theta_n$,
and where the set $\Delta$ contains all the free variables appearing in
$\Gamma, M,\theta$. 

\begin{figure*}
\begin{frameit}
$$
\begin{array}{@{}c@{}}
\prooftree
\,
\justifies
\Gamma, x\colon \theta \vdash_\Delta x : \theta
\endprooftree
\qquad
\prooftree
\Gamma, x\colon \theta \vdash_\Delta M : \theta'
\justifies
\Gamma \vdash_\Delta \lambda x\colon \theta. M : \theta\rightarrow\theta'
\endprooftree
\qquad
\prooftree
\Gamma \vdash_\Delta M : \theta'\rightarrow\theta
\quad
\Gamma \vdash_\Delta M' : \theta'
\justifies
\Gamma \vdash_\Delta M M' : \theta
\endprooftree
\\
\\
\prooftree
\Gamma \vdash_{\Delta\cup \{i\}} M : \theta'
\justifies
\Gamma \vdash_\Delta \lambda i. M : \Pi_i\theta'
\using
(i\not\in\FIV(\Gamma,\Delta))
\endprooftree
\qquad
\prooftree
\Gamma \vdash_\Delta M : \Pi_i\theta
\quad
\Delta \vdash E
\justifies
\Gamma \vdash_\Delta M E : \theta[E/i]
\endprooftree
\qquad
\prooftree
\Gamma \vdash_\Delta M : \theta \rightarrow \theta
\justifies
\Gamma \vdash_\Delta \fix\; M : \theta
\endprooftree
\\
\\
\prooftree
\Gamma \vdash_\Delta M : \trity{P\wedge E{=}0}{Q}
\quad
\Gamma \vdash_\Delta M' : \trity{P\wedge E{\not=}0}{Q}
\justifies
\Gamma \vdash_\Delta \ifz\,E\,M\,M' : \trity{P}{Q}
\endprooftree
\qquad
\prooftree
\Delta \vdash P
\justifies
\Gamma \vdash_\Delta \myskip : \trity{P}{P}
\endprooftree
\\
\\
\prooftree
\Gamma \vdash_\Delta M : \trity{P}{P'}
\quad
\Gamma \vdash_\Delta M' : \trity{P'}{Q}
\justifies
\Gamma \vdash_\Delta (M;M') : \trity{P}{Q}
\endprooftree
\qquad
\prooftree
\Delta \vdash E
\justifies
\Gamma \vdash_\Delta \free(E) : \trity{E{\mapsto} {-}}{\emp}
\endprooftree
\\
\\
\prooftree
\Gamma \vdash_{\Delta\cup\{i\}}
  M :
  \trity{i{\mapsto}{-} * P}{Q}
\justifies
\Gamma \vdash_\Delta
  \mylet{i=\new}{M} :
  \trity{P}{Q}
\using
(i \not\in \FIV(\Gamma,\Delta,P,Q))
\endprooftree
\qquad
\prooftree
\Delta \vdash E 
\quad
\Delta \vdash E'
\justifies
\Gamma \vdash_\Delta [E]{:=}E' : \trity{E{\mapsto} -}{E{\mapsto} E'}
\endprooftree
\\
\\
\prooftree
\Gamma \vdash_{\Delta\cup\{i\}} M : \trity{E{\mapsto}i*P}{Q}
\justifies
\Gamma \vdash_\Delta
   \mylet{i{=}[E]}{M} : \trity{\exists i.\,E{\mapsto}i*P}{Q}
\using
(i \not\in \FIV(\Gamma,\Delta,E,Q))
\endprooftree
\qquad
\prooftree
\Gamma \vdash_\Delta M : \theta
\quad
\theta \preceq_\Delta \theta'
\justifies
\Gamma \vdash_\Delta M : \theta'
\endprooftree
\end{array}
$$
\caption{Typing Rules}
\label{fig:typing-rules}
\end{frameit}
\end{figure*}

\begin{figure*}
\begin{frameit}
\begin{center}
\mbox{\sc Inference Rules}
\end{center}
$$
\begin{array}{c}
\\
\prooftree
  \theta \preceq_\Delta \theta
  \quad
  \theta' \preceq_\Delta \theta''
\justifies
  \theta \preceq_\Delta \theta''
\endprooftree
\qquad
\prooftree
  \theta'_0 \preceq_\Delta \theta_0
  \quad
  \theta_1 \preceq_\Delta \theta_1'
\justifies
  (\theta_0 \rightarrow \theta_1) 
  \preceq_\Delta 
  (\theta'_0 \rightarrow \theta'_1) 
\endprooftree
\qquad
\prooftree
  \theta \preceq_{\Delta \cup \{i\}} \theta'
\using
  (i\not\in \Delta)
\justifies
  (\Pi_i \theta) \preceq_\Delta (\Pi_i \theta')
\endprooftree
\\
\\
\prooftree
  \theta \preceq_\Delta \theta'
\justifies
  (\theta \otimes P) \preceq_\Delta (\theta' \otimes P)
\endprooftree
\qquad
\prooftree
   \forall \eta \in \ff{\Delta}.
   \, \ff{P'}_\eta \subseteq \ff{P}_\eta 
      \wwedge
      \ff{Q}_\eta \subseteq \ff{Q'}_\eta
\justifies
   \trity{P}{Q} \preceq_\Delta \trity{P'}{Q'} 
\endprooftree
\\
\\[1ex]
\end{array}
$$

\begin{center}
\mbox{\sc Axioms}
\end{center}
$$
\begin{array}{@{}r@{\;}c@{\;}l@{\;\;\;}r@{\;}c@{\;}l@{\;\;\;}r@{\;}c@{\;}l@{}}
\\
  \theta & \preceq_\Delta & \theta
&
  \theta 
  & \preceq_\Delta & 
  \theta \otimes P
\\
  (\trity{P}{Q})\otimes P_0 
  & \simeq_\Delta & 
  \trity{P*P_0}{Q*P_0}
&
  (\Pi_i \theta)\otimes P 
  & \simeq_\Delta &
  \Pi_i (\theta\otimes P)
  \;\;(\mbox{when } i{\not\in} \Delta)
\\
  (\theta\otimes Q)\otimes P 
  & \simeq_\Delta & 
  \theta\otimes (Q*P)
&
  (\theta \rightarrow \theta')\otimes P 
  & \simeq_\Delta &
  (\theta\otimes P \rightarrow \theta'\otimes P) 
\end{array}
$$
\caption{Rules for the Subtyping Relation $\preceq_\Delta$}
\label{fig:subtyping-rules}
\end{frameit}
\end{figure*}

The type system is shown in Figures~\ref{fig:typing-rules}
and \ref{fig:subtyping-rules}.  For
notational simplicity we have omitted some obvious side-conditions of
the form $\Delta\vdash\theta:\type$ which ensure that, for a judgment
$\Gamma \vdash_\Delta M : \theta$, the set $\Delta$ always
contains all the free variables appearing in $\Gamma, M,\theta$, and
that the type assignment $\Gamma$ is always well-formed.  There are
three classes of rules. The first class consists of the rules from the
simply typed lambda calculus extended with dependent product types and
recursion. The second class consists of the rules for the imperative
constructs, all of which come from separation logic. The last class
consists of the subsumption rule based on the subtype relation
$\preceq_\Delta$, which is the most interesting part of our type
system. The proof rules for $\preceq_\Delta$ are shown in 
Figure~\ref{fig:subtyping-rules}. They define a preorder between
types with free variables in $\Delta$, and include all the usual
structural subtyping rules in the chapter 15 of \cite{pierce-types-prog-lang-02}.
The rules specific to our system are: the covariant structural rule for
$\theta \otimes P$; the encoding of Consequence in
Hoare logic; the generalized frame rule that adds an invariant to all
types; and the distribution rules for an added invariant assertion.

The generalized frame rule, $\theta \preceq_\Delta \theta\otimes P_0$,
means that if a program satisfies $\theta$ and an assertion $P_0$ does
not ``mention'' any cells described by $\theta$, then the program
preserves $P_0$. Note that this rule indicates that the types in our
system are \emph{tight}~\cite{ohearn-ishtiaq-popl01,reynolds02}: if a
program satisfies $\theta$, it can only access heap cells
``mentioned'' in $\theta$. This is why an assertion $P_0$ for
``unmentioned'' cells is preserved by the program. For instance, if a
program $M$ has a type of the form
$$
\theta_1 \rightarrow \ldots \rightarrow \theta_n
\rightarrow \trity{P}{Q},
$$
the tightness of the type says that all the cells that $M$ can 
directly access must appear in the pre-condition $P$. Thus,
if no cells in an assertion $P_0$ appear in $P$, program $M$ maintains
$P_0$, as long as argument procedures maintain it. 
Such a fact can, indeed, be inferred by the generalized frame rule together 
with the distribution rules:
$$
\begin{array}{l}
\qquad
  \theta_1 \rightarrow \ldots \rightarrow \theta_n
  \rightarrow \trity{P}{Q} 
\\
\preceq_\Delta 
\qquad (\because~  \theta \preceq_\Delta \theta \otimes P_0)
\\
\qquad
  (\theta_1 \rightarrow \ldots \rightarrow \theta_n
   \rightarrow \trity{P}{Q}) \otimes P_0
\\
\preceq_\Delta
\qquad (\because~  (\theta \rightarrow \theta') \otimes P_0
                \preceq_\Delta 
                (\theta\otimes P_0 \rightarrow \theta' \otimes P_0))
\\
\qquad
  (\theta_1 \otimes P_0 \rightarrow \ldots \rightarrow \theta_n \otimes P_0
  \rightarrow \trity{P}{Q} \otimes P_0)
\\
\preceq_\Delta
\qquad (\because~  \trity{P}{Q} \otimes P_0
                \preceq_\Delta 
                \trity{P*P_0}{Q*P_0})
\\
\qquad
  (\theta_1 \otimes P_0 \rightarrow \ldots \rightarrow \theta_n \otimes P_0
  \rightarrow \trity{P*P_0}{Q*P_0}).
\end{array}
$$
The generalized frame rule, the distribution rules, and
the structural subtyping rule for function types all together give many
interesting higher-order frame rules, including the second-order frame rule.
The common mechanism for obtaining such a rule is:
first, add an invariant assertion by the generalized frame rule,
and then, propagate the added assertion all the way down to
a base triple type by the distribution rules.
The structural subtyping rule for the function type allows us
to apply this construction for a sub type-expression in 
an appropriate covariant or contravariant way. For instance,
we can derive a third-order frame rule as follows:
$$
\begin{array}{@{}l@{}}
\qquad
(\trity{P_1}{Q_1}\rightarrow\trity{P_2}{Q_2}) \rightarrow \trity{P_3}{Q_3}
\\
\preceq_\Delta \qquad (\because~ \theta \preceq_\Delta \theta \otimes P)
\\
\qquad
\Bigl((\trity{P_1}{Q_1}\rightarrow\trity{P_2}{Q_2}) 
      \rightarrow \trity{P_3}{Q_3}\Bigr) \otimes P
\\
\preceq_\Delta
\qquad (\because~ (\theta \rightarrow \theta') \otimes P
                \simeq_\Delta 
                (\theta\otimes P \rightarrow \theta' \otimes P))
\\
\qquad
(\trity{P_1}{Q_1}{\otimes} P\rightarrow\trity{P_2}{Q_2} {\otimes} P)
      \rightarrow \trity{P_3}{Q_3}{\otimes} P
\\
\preceq_\Delta
\qquad (\because~ \mbox{structural subtyping})
\\
\qquad
(\trity{P_1}{Q_1}{\otimes}P\rightarrow\trity{P_2}{Q_2})
      \rightarrow \trity{P_3}{Q_3}{\otimes} P
\\
\preceq_\Delta
\qquad (\because~ \trity{P_0}{Q_0} \otimes P
                \simeq_\Delta 
                \trity{P_0*P}{Q_0*P})
\\
\qquad
(\trity{P_1{*}P}{Q_1{*}P}\rightarrow\trity{P_2}{Q_2})
      \rightarrow \trity{P_3{*}P}{Q_3{*}P}.
\end{array}
$$

\subsection{Example Proofs in the Type System}\label{sec:example-proofs}
We illustrate how the type system works, with the verification
of two example programs. 

The first example is a procedure that disposes a linked list.
With this example we demonstrate how a standard proof in separation logic
yields a typing in our type system.  Let $\lst(i)$ be an assertion which expresses
that the heap contains a linked 
list $i$ terminating with $0$, and all the cells in the heap are 
in the list.\footnote{Formally, $\lst(i)$ is the (parameterized)
assertion that satisfies the equivalence:
$$
   \lst(i) \iff
   (i=0 \wedge \emp) \vee
   (\exists j.\, (i \mapsto j)*\lst(j))
$$ --- it can be defined as the minimal fixed point, expressible in
higher-order separation logic~\cite{biering-birkedal-torpsmith-esop05}.
} 
We define a procedure $\dlist$ for list disposal as follows:
$$
  \dlist 
  \;\defeq\;
  \fix\; \lambda f\colon (\Pi_i\trity{\lst(i)}{\emp}).\,
         \Bigl(\lambda i.\; \ifz\; i\; \bigl(\myskip\bigr)\; 
                                 (\mylet{j=[i]}{f(j);\free(i)})\Bigr).
$$
The program $\dlist$ takes a linked list $i$, and disposes
the list, first the tail and then the head of the list.

We derive the typing judgment 
${} \vdash_{\{\}} \dlist : (\Pi_i\trity{\lst(i)}{\emp})$.
Note that this derivation captures the correctness of $\dlist$,
because the judgment means that when $\dlist$ is given 
a linked list $i$ as argument, then it disposes all the cells in the list.

The main part of the derivation is a proof tree 
for the false branch of the conditional
statement. Let $\Gamma$ be $f\colon (\Pi_i\trity{\lst(i)}{\emp})$.
The proof tree for the false branch is given below:
$$
\prooftree
  \prooftree
    \prooftree
      \prooftree
        \prooftree
          \prooftree
          \justifies
            \Gamma \vdash_{\{i,j\}}
            f :
            \Pi_i\trity{\lst(i)}{\emp} 
          \endprooftree
        \justifies
            \Gamma \vdash_{\{i,j\}}
            f(j) :
            \trity{\lst(j)}{\emp}
        \endprooftree
      \using
         {1}
      \justifies
         \Gamma \vdash_{\{i,j\}}
         f(j) :
         \trity{i{\mapsto}j *\lst(j)}{i{\mapsto}j}
      \endprooftree
      \quad
      \prooftree
         \prooftree
         \justifies
           \Gamma \vdash_{\{i,j\}}
           \free(i) :
           \trity{i{\mapsto} -}{\emp}
         \endprooftree
      \using
         2
      \justifies
         \Gamma \vdash_{\{i,j\}}
         \free(i) :
         \trity{i{\mapsto} j}{\emp}
      \endprooftree
    \justifies
      \Gamma \vdash_{\{i,j\}}
      (f(j);\free(i)) :
      \trity{i{\mapsto}j *\lst(j)}{\emp}
    \endprooftree
  \justifies
    \Gamma \vdash_{\{i\}}
    (\mylet{j{=}[i]}{f(j);\free(i)}) :
    \trity{\exists j.\, {i{\mapsto} j}*\lst(j)}{\emp}
  \endprooftree
\using
  3
\justifies
  \Gamma \vdash_{\{i\}}
  (\mylet{j{=}[i]}{f(j);\free(i)}) :
  \trity{\lst(i) \wedge i{\not=}0}{\emp}
\endprooftree
$$
Most of the steps in this tree use syntax-directed
rules, such as those for the sequential composition and procedure
application. The only exceptions are the steps marked by
$1$, $2$ and $3$, where we apply the subsumption rule.
These steps express structural rules in separation logic.
Step $1$ is an instance of the ordinary frame rule,
and attaches the invariant $(i{\mapsto} j)$ to the pre- 
and post-conditions of the triple type $\trity{\lst(j)}{\emp}$.
The other steps are an instance of Consequence.
Step $2$ strengthens the pre-condition
of $\trity{i{\mapsto}{-}}{\emp}$,
and step $3$ replaces the pre-condition of 
$\trity{\exists j.i{\mapsto}j *\lst(j)}{\emp}$ by the equivalent
assertion $\lst(j) \wedge i{\not=}0$. 
In the tree above, 
we have not shown how to derive the necessary subtype relations in $1$,
$2$ and $3$.  They are straightforward to derive:
$$
\begin{array}{@{}l@{}}
\begin{array}{@{}r@{\,}c@{\,}l@{}}
\trity{\lst(j)}{\emp} 
& \preceq_{\{i,j\}} & (\trity{\lst(j)}{\emp}) \otimes {i{\mapsto}j} 
\;\hfill
(\because~ \theta \;\preceq_\Delta\; \theta \otimes R)
\\
& \preceq_{\{i,j\}} & \trity{\lst(j)*i{\mapsto}j}{\emp * i{\mapsto}j} 
\;\hfill
(\because~ \trity{P}{Q}\otimes R \;\simeq_\Delta\; \trity{P*R}{Q*R})
\\
& \preceq_{\{i,j\}} & \trity{i{\mapsto}j * \lst(j)}{i{\mapsto}j}
\;\hfill
(\because~ \forall \eta.\; 
           \ff{P {*} Q}_\eta = \ff{Q {*} P}_\eta \wwedge
           \ff{\emp {*} P}_\eta = \ff{P}_\eta)
\end{array}
\\
\\
\begin{array}{@{}r@{\,}c@{\,}l@{}}
\trity{i{\mapsto}{-}}{\emp} 
& \preceq_{\{i,j\}} & 
\trity{i{\mapsto}j}{\emp} 
\;\hfill
(\because~ \forall \eta.\;
           \ff{i{\mapsto}j}_\eta \subseteq
           \ff{i{\mapsto}-}_\eta) 
\end{array}
\\
\\
\begin{array}{@{}r@{\,}c@{\,}l@{}}
\trity{\exists j.\,i{\mapsto}j * \lst(j)}{\emp} 
& \preceq_{\{i\}} & 
\trity{\lst(i) \wedge i{\not=}0}{\emp} 
\,\hfill
(\because \forall \eta.\,
           \ff{\exists j.i{\mapsto}j {*} \lst(j)}_\eta \,{=}\, 
           \ff{\lst(i) \wedge i{\not=}0}_\eta).
\end{array}
\end{array}
$$
The complete derivation of ${} \vdash_{\{\}} \dlist : \Pi_i\trity{\lst(i)}{\emp}$ is shown in Figure~\ref{fig:derivation-dlist}. 

\begin{figure*}
\begin{frameit}
{\small
$$
\prooftree
  \prooftree
    \prooftree
      \prooftree
        \prooftree
          \prooftree
            \prooftree
              \prooftree
                \prooftree
                  \prooftree
                  \justifies
                    \Gamma \vdash_{\{i,j\}}
                    f {:}
                    \Pi_i\trity{\lst(i)}{\emp} 
                  \endprooftree
                \justifies
                    \Gamma \vdash_{\{i,j\}}
                    f(j) {:}
                    \trity{\lst(j)}{\emp}
                \endprooftree
              \using
                 {1}
              \justifies
                 \Gamma \vdash_{\{i,j\}}
                 f(j) {:}
                 \trity{i{\mapsto}j *\lst(j)}{i{\mapsto}j}
              \endprooftree
              \!\!
              \prooftree
                 \prooftree
                 \justifies
                   \Gamma \vdash_{\{i,j\}}
                   \free(i) {:}
                   \trity{i{\mapsto}{-}}{\emp}
                 \endprooftree
              \using
                 2
              \justifies
                 \Gamma \vdash_{\{i,j\}}
                 \free(i) {:}
                 \trity{i{\mapsto} j}{\emp}
              \endprooftree
            \justifies
              \Gamma \vdash_{\{i,j\}}
              (f(j);\free(i)) {:}
              \trity{i{\mapsto}j *\lst(j)}{\emp}
            \endprooftree
          \justifies
            \Gamma \vdash_{\{i\}}
            \mylet{j{=}[i]}{(f(j);\free(i))} {:}
            \trity{\exists j.\, {i{\mapsto} j}*\lst(j)}{\emp}
          \endprooftree
        \using
          {3}
        \justifies
          \Gamma \vdash_{\{i\}}
          \mylet{j{=}[i]}{(f(j);\free(i))} {:}
          \trity{\lst(i) \wedge i{\not=}0}{\emp}
        \endprooftree
        \!\!\!\!\!\!\!\!\!\!\!\!\!
        \prooftree
          \prooftree
          \justifies
            \Gamma \vdash_{\{i\}} \myskip {:} \trity{\emp}{\emp}
          \endprooftree
        \using
          {4}
        \justifies
          \Gamma \vdash_{\{i\}} \myskip {:} \trity{\lst(i) \wedge i{=}0}{\emp}
        \endprooftree
      \justifies
        \Gamma \vdash_{\{i\}}
        \bigl(\ifz\;i\;(\myskip)\;(\mylet{j{=}[i]}{(f(j);\free(i))})\bigr) {:}
        \trity{\lst(i)}{\emp}
      \endprooftree
    \justifies
      \Gamma \vdash_{\{\}}
      \bigl(\lambda i.\,
      \ifz\;i\;(\myskip)\;(\mylet{j{=}[i]}{(f(j);\free(i))})\bigr) {:}
      \Pi_i\trity{\lst(i)}{\emp}
    \endprooftree
  \justifies
  {} \vdash_{\{\}}
    \bigl(
    \lambda f.
    \lambda i.\,
    \ifz\;i\;(\myskip)\;(\mylet{j{=}[i]}{(f(j);\free(i))})\bigr) {:}
    (\Pi_i\trity{\lst(i)}{\emp}) 
    \rightarrow 
    (\Pi_i\trity{\lst(i)}{\emp})
  \endprooftree
\justifies
  {} \vdash_{\{\}}
    \dlist {:}  \Pi_i\trity{\lst(i)}{\emp}
\endprooftree
$$
}
In the tree, $\Gamma$ is $f\colon \Pi_i\trity{\lst(i)}{\emp}$;
and at $1$ - $4$ of the tree, 
the subsumption rule is used with the following subtype relations:
$$
\begin{array}{@{}r@{\,}c@{\,}l@{}}
\trity{\lst(j)}{\emp} 
& \preceq_{\{i,j\}} & 
\trity{i{\mapsto}j * \lst(j)}{i{\mapsto}j}
\\
\trity{i{\mapsto}{-}}{\emp} 
& \preceq_{\{i,j\}} & 
\trity{i{\mapsto}j}{\emp} 
\\
\trity{\exists j.\,i{\mapsto}j * \lst(j)}{\emp} 
& \preceq_{\{i\}} & 
\trity{\lst(i) \wedge i{\not=}0}{\emp} 
\\
\trity{\emp}{\emp}
& \preceq_{\{i\}} & 
\trity{\lst(i) \wedge i{=}0}{\emp} 
\end{array}
$$
\caption{Derivation of the Typing Judgment 
         ${} \vdash_{\{\}} \dlist : \Pi_i\trity{\lst(i)}{\emp}$}
\label{fig:derivation-dlist}
\end{frameit}
\end{figure*}

The second example is a client program that uses a randomized memory 
manager. The verification of this program demonstrates the use 
of a third-order frame rule. 

The randomized memory manager is a module with two methods, 
$\Malloc$ for allocating a cell and $\Mfree$ for deallocating a cell.
The memory manager maintains a free list whose starting address 
is stored in the cell $l$. When $\Malloc$ is called, the module first 
checks this free list $[l]$. If the free list is not empty, $\Malloc$ 
takes one cell from the list and returns it to the client. Otherwise, 
$\Malloc$ makes a system call, obtains a new cell from the operating 
system, and returns this cell to the client. 
When $\Mfree$ is called to deallocate cell $i$,
the randomized memory manager first flips a coin. Then,
depending on the result of the coin, it either adds the cell $i$ 
to the free list or returns the cell to the operating system. 
Note that randomization is used only in $\Mfree$. 
We will focus on the method $\Mfree$.

Let $\inv(l)$ be the assertion $\exists l'.\,(l{\mapsto}l') * \lst(l')$,
which expresses that cell $l$ stores the starting address of a linked list.
The following program implements the $\Mfree$ method of 
the randomized memory manager:
$$
\begin{array}{rcl}
  \Mfree & : & \Bigl(\bigl(\Pi_i 
                  \trity
                    {{i{\mapsto}{-}}}
                    {{i{\mapsto}{-}}}\bigr)
               \rightarrow
               \bigl(\Pi_i
                  \trity
                    {{i{\mapsto}{-}}}
                    {\emp}\bigr)
               \Bigr)
               \otimes \inv(l) 
\\
  \Mfree
         & \defeq & \lambda \coinflip.\; \lambda i.\;
                       \coinflip(i);
                       \mylet
                       {i'{=}[i]}
                       {\Bigl(\ifz\;i'\;
                            (\free(i))\;
                            \bigl(\mylet{l'{=}[l]}{([i]{:=}l';[l]{:=}i)}\bigr)
                        \Bigr)}
\end{array}
$$
Note that before disposing cell $i$,
method $\Mfree$ uses the cell to store the result of flipping a coin by
calling $\coinflip$ with $i$. The declared type of the method $\Mfree$ 
has the form $\theta \otimes \inv(l)$. The $\theta$ part expresses
that the method has the expected behavior externally, and
the $\inv(l)$ part indicates that it maintains the module invariant 
internally. The derivation of the declared type is shown in 
Figure~\ref{fig:derivation-free}.

\begin{figure*}
\begin{frameit}
Consider $\Gamma,\Delta$ such that $\coinflip \not\in \dom(\Gamma)$
and $l',i',i \not\in \Delta$ but $l \in \Delta$.
Define $\Gamma'$, $\FalseBranch$, and $\Body$ as follows:
$$
\begin{array}{rcl}
   \Gamma' & \defeq & \Gamma,\, \coinflip\colon
                  \Pi_i
                  \trity
                    {{i{\mapsto}{-}}}
                    {{i{\mapsto}{-}}}
\\
   \FalseBranch & \defeq & \mylet{l'{=}[l]}{([i]{:=}l';[l]{:=}i)}
\\
   \Body & \defeq & 
           \mylet{i'{=}[i]}{(\ifz\;i'\;
                            (\free(i))\;
                            \FalseBranch)}
\end{array}
$$
The term $\Mfree$ and its subterms $\FalseBranch$ and $\Body$ 
are typed as follows:
{\small
$$
\prooftree
  \prooftree
    \prooftree
      \prooftree
        \prooftree
          \prooftree
          \justifies
            \Gamma' \vdash_{\Delta \cup \{i,i',l'\}} 
                        [i]{:=}l'
             {:} \trity{i{\mapsto}{-}}{i{\mapsto} l'}
          \endprooftree
        \using
          1
        \justifies
          \Gamma' \vdash_{\Delta \cup \{i,i',l'\}} 
                      [i]{:=}l'
           {:} \trity{l{\mapsto}{-} {*} i{\mapsto}{-}}
                   {l{\mapsto}{-} {*} i{\mapsto} l'}
        \endprooftree
        \prooftree
          \prooftree
          \justifies
            \Gamma' \vdash_{\Delta \cup \{i,i',l'\}} 
                        [l]{:=}i
             {:} \trity{l{\mapsto}{-}}{l{\mapsto}i}
          \endprooftree
        \using
          2
        \justifies
          \Gamma' \vdash_{\Delta \cup \{i,i',l'\}} 
                      [l]{:=}i
           {:} \trity{l{\mapsto}{-} {*} i{\mapsto}l'}
                   {l{\mapsto}i {*} i{\mapsto} l'}
        \endprooftree
      \justifies
        \Gamma' \vdash_{\Delta \cup \{i,i',l'\}} 
                     ([i]{:=}l';[l]{:=}i)
          {:} \trity{l{\mapsto}{-} {*} i{\mapsto}{-}}
                  {l{\mapsto}i {*} i{\mapsto} l'}
      \endprooftree
    \using
       3
    \justifies
       \Gamma' \vdash_{\Delta \cup \{i,i',l'\}} 
                     ([i]{:=}l';[l]{:=}i)
          {:} \trity{l{\mapsto}l' {*} i{\mapsto}i'{*}\lst(l')}{l{\mapsto}i {*} i{\mapsto}l' {*} \lst(l')}
    \endprooftree
  \justifies
     \Gamma' \vdash_{\Delta \cup \{i,i'\}} 
                   \mylet{l'{=}[l]}{([i]{:=}l';[l]{:=}i)}
        {:} \trity{\exists l'.\, l{\mapsto}l' {*} i{\mapsto}i'{*}\lst(l')}{l{\mapsto}i {*} i{\mapsto}l' {*} \lst(l')}
  \endprooftree
\using
  4
\justifies
  \Gamma' \vdash_{\Delta \cup \{i,i'\}} 
     \FalseBranch
     {:} \trity{{i{\mapsto}i'{*}\inv(l) \wedge i'{\not=}0}}{\inv(l)}
\endprooftree
$$
$$
\prooftree
  \prooftree
    \prooftree
      \prooftree
        \prooftree
        \justifies
          \Gamma' \vdash_{\Delta{\cup}\{i,i'\}} 
             \free(i)
             {:} \trity{i{\mapsto}{-}}{\emp}
        \endprooftree
      \using
        5
      \justifies
        \Gamma' \vdash_{\Delta {\cup} \{i,i'\}} 
           \free(i)
           {:} \trity{{i{\mapsto}i'{*}\inv(l) \wedge i'{=}0}}{\inv(l)}
      \endprooftree
      %
      %
        \Gamma' \,{\vdash_{\Delta {\cup} \{i,i'\}}} 
                      \FalseBranch
           {:} \trity{{i{\mapsto}i'{*}\inv(l) \wedge i'{\not=}0}}{\inv(l)}
    \justifies
      \Gamma' \vdash_{\Delta \cup \{i,i'\}} 
          (\ifz\;i'\; (\free(i))\; \FalseBranch)
         {:} \trity{{i{\mapsto}i'{*}\inv(l)}}{\inv(l)}
    \endprooftree
  \justifies
    \Gamma' \vdash_{\Delta \cup \{i\}} 
        \mylet{i'{=}[i]}{(\ifz\;i'\;
                            (\free(i))\;
                            \FalseBranch)}
       {:} \trity{{\exists i'.i{\mapsto}i'{*}\inv(l)}}{\inv(l)}
  \endprooftree
\using
  6
\justifies
  \Gamma' \vdash_{\Delta \cup \{i\}} 
     \Body
     {:} \trity{{i{\mapsto}{-}{*}\inv(l)}}{\inv(l)}
\endprooftree
$$
$$
\prooftree
  \prooftree
    \prooftree
      \prooftree
        \prooftree
          \prooftree
          \justifies
             \Gamma' \vdash_{\Delta \cup \{i\}}
                \coinflip(i) {:} \trity{i{\mapsto}{-}}
                                     {i{\mapsto}{-}}
          \endprooftree
        \using
          7
        \justifies
          \Gamma' \vdash_{\Delta \cup \{i\}}
             \coinflip(i) {:} \trity{i{\mapsto}{-}{*}\inv(l)}
                                  {i{\mapsto}{-}{*}\inv(l)}
        \endprooftree
        \quad
        %
          \Gamma' \vdash_{\Delta \cup \{i\}} 
             \Body
             {:} \trity{{i{\mapsto}{-}{*}\inv(l)}}{\inv(l)}
      \justifies
        \Gamma' \vdash_{\Delta \cup \{i\}}
           \coinflip(i); \Body
           {:}
           \trity{{i{\mapsto}{-}{*}\inv(l)}}{\inv(l)}
      \endprooftree
    \justifies
      \Gamma' \vdash_\Delta
      \lambda i.\;
         \coinflip(i);
         \Body
         {:}
         \Pi_i\trity{{i{\mapsto}{-}{*}\inv(l)}}{\inv(l)}
    \endprooftree
  \justifies
    \Gamma \vdash_\Delta 
    \Mfree {:}   \bigl(\Pi_i 
                    \trity
                      {{i{\mapsto}{-}}{*}\inv(l)}
                      {{i{\mapsto}{-}}{*}\inv(l)}\bigr)
                 \rightarrow
                 \bigl(\Pi_i
                    \trity
                      {{i{\mapsto}{-}{*}\inv(l)}}
                      {\inv(l)}\bigr)
  \endprooftree
\using
  8
\justifies
  \Gamma \vdash_\Delta
  \Mfree {:}     \Bigl(\bigl(\Pi_i 
                  \trity
                    {{i{\mapsto}{-}}}
                    {{i{\mapsto}{-}}}\bigr)
               \rightarrow
               \bigl(\Pi_i
                  \trity
                    {{i{\mapsto}{-}}}
                    {\emp}\bigr)
               \Bigr)
               \otimes \inv(l) 
\endprooftree
$$
}
where the steps marked by 1-8 use the subsumption rule.
\caption{Derivation of the Typing Judgment for $\Mfree$}
\label{fig:derivation-free}
\end{frameit}
\end{figure*}

We now consider the following client of the randomized memory manager.
$$
\begin{array}{rcl}
\Rand & : & \Pi_i \trity
                     {i{\mapsto}{-}}
                     {i{\mapsto}{-}}
\\
\Rand & \defeq & \lambda i.\;
                   \mylet
                    {i'=[i]}
                     {[i]{:=}i'{+}1}
\\
\\
\Client & : & (\Pi_i 
                  \trity
                    {{i{\mapsto}{-}}}
                    {{i{\mapsto}{-}}}
               \rightarrow
               \Pi_i
                  \trity
                    {{i{\mapsto}{-}}}
                    {\emp})
               \rightarrow
               \trity{j{\mapsto}{-}}
                     {\emp}
\\
\Client & \defeq & \lambda \mfree.\;
                   (\mfree\;\Rand\;j)
\end{array}
$$
The client $\Client$ takes a ``randomized'' method $\mfree$ 
for deallocating a cell. Then, it instantiates the method with the 
(degenerate) ``random function'' $\Rand$,
and calls the instantiated method to dispose cell $j$. Suppose
that $\Client$ is ``linked'' with the $\Mfree$ of the randomized memory
manager, that is, that it is applied to $\Mfree$. We prove the
correctness of this application by deriving the typing judgment
$\vdash_{\{j,l\}} (\Client\;\Mfree) : \trity{j{\mapsto}{-}*\inv(l)}{\inv(l)}$.

The derivation of the mentioned typing judgment consists
of three parts: the sub proof-trees for $\Mfree$ and $\Client$, 
and the part that links these two proof trees. The sub proof-trees
for $\Mfree$ and $\Client$ are shown in Figures~\ref{fig:derivation-free} 
and \ref{fig:derivation-client}. Note that the internal 
free list $[l]$ of the memory manager does not appear in the proof 
tree for $\Client$ in Figure~\ref{fig:derivation-client}; all the 
rules in the tree concern just cell $j$, the only cell that $\Client$ 
directly manipulates.

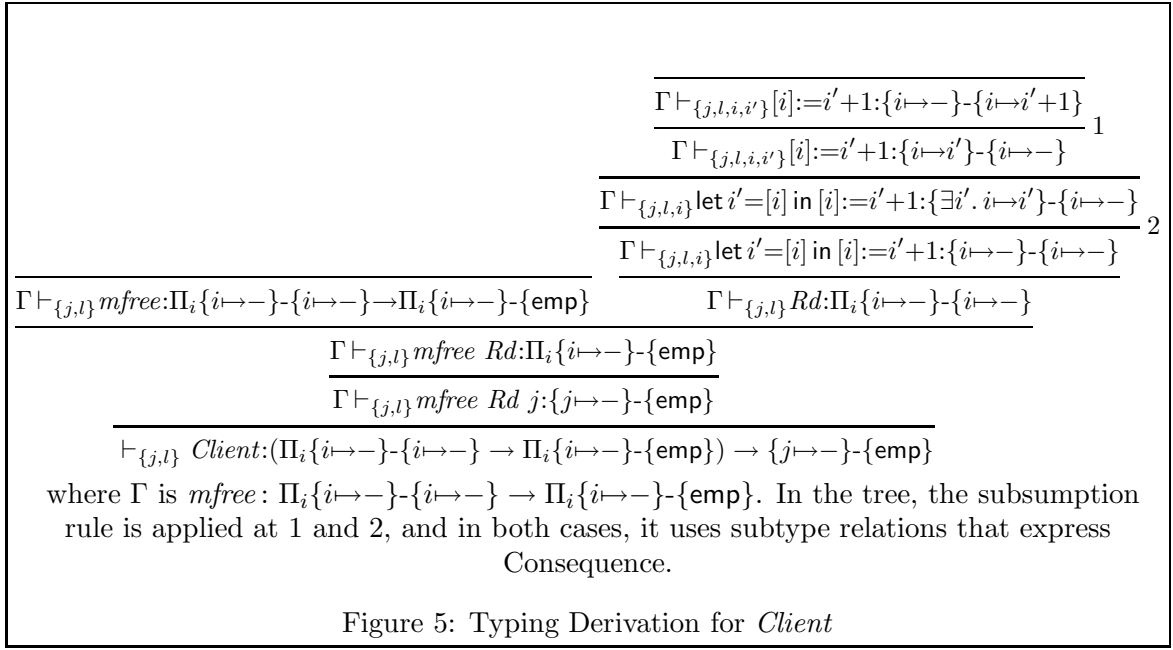
\begin{figure*}
\begin{frameit}
{\small
$$
  \prooftree
    \prooftree
      \prooftree
          \prooftree
          \justifies
            \Gamma
            \,{\vdash_{\{j,l\}}}
            \mfree{:} \Pi_i 
                      \trity
                        {i{\mapsto}{-}}
                        {i{\mapsto}{-}}
                     {\rightarrow}
                     \Pi_i
                       \trity
                        {i{\mapsto}{-}}
                        {\emp}
          \endprooftree
        \!\!\!\!
        \prooftree
          \prooftree
            \prooftree
              \prooftree
                \prooftree
                \justifies
                  \Gamma 
                  \,{\vdash_{\{j,l,i,i'\}}}
                         [i]{:=}i'{+}1
                  {:} \trity
                      {i{\mapsto}{-}}
                      {i{\mapsto}{i'{+}1}}
                \endprooftree
              \using
                1
              \justifies
                \Gamma 
                \,{\vdash_{\{j,l,i,i'\}}}
                       [i]{:=}i'{+}1
                {:} \trity
                    {i{\mapsto}{i'}}
                    {i{\mapsto}{-}}
              \endprooftree
            \justifies
              \Gamma 
              \,{\vdash_{\{j,l,i\}}}
                    \mylet
                     {i'{=}[i]}
                     {[i]{:=}i'{+}1}
              {:} \trity
                  {\exists i'.\, i{\mapsto}{i'}}
                  {i{\mapsto}{-}}
            \endprooftree
          \using
            2 
          \justifies
            \Gamma 
            \,{\vdash_{\{j,l,i\}}}
                  \mylet
                   {i'{=}[i]}
                   {[i]{:=}i'{+}1}
            {:} \trity
                {i{\mapsto}{-}}
                {i{\mapsto}{-}}
          \endprooftree
        \justifies
          \Gamma 
          \,{\vdash_{\{j,l\}}}
          \Rand {:} \Pi_i \trity
                   {i{\mapsto}{-}}
                   {i{\mapsto}{-}}
        \endprooftree
      \justifies
        \Gamma
        \,{\vdash_{\{j,l\}}}
        \mfree\;\Rand {:} \Pi_i \trity{i{\mapsto}{-}}
                                    {\emp}
      \endprooftree
    \justifies
      \Gamma
      \,{\vdash_{\{j,l\}}}
      \mfree\;\Rand\;j {:} \trity{j{\mapsto}{-}}
                               {\emp}
    \endprooftree
  \justifies
    {} \vdash_{\{j,l\}}
         \Client  {:}  {(
                          \Pi_i 
                                \trity
                                  {{i{\mapsto}{-}}}
                                  {{i{\mapsto}{-}}}
                          \rightarrow
                          \Pi_i
                                \trity
                                  {{i{\mapsto}{-}}}
                                  {\emp}
                        )} 
                       \rightarrow
                       \trity{j{\mapsto}{-}}
                             {\emp}
  \endprooftree
$$
}
where $\Gamma$ is 
$\mfree \colon 
          \Pi_i 
             \trity{i{\mapsto}{-}}{i{\mapsto}{-}}
          \rightarrow
          \Pi_i 
              \trity{i{\mapsto}{-}}{\emp}$.
In the tree, the subsumption rule is applied at $1$ and $2$, 
and in both cases, it uses subtype relations that express 
Consequence.
\caption{Typing Derivation for $\Client$}
\label{fig:derivation-client}
\end{frameit}
\end{figure*}

It is the third-order frame rule that lets us ignore 
the internal free list $[l]$ of the memory manager when
constructing the proof tree for $\Client$.
The third-order frame rule adds
the missing free list $[l]$ to the derived type for $\Client$,
so that we can link $\Client$ with $\Mfree$, without producing
a type error. More precisely, the rule allows the following
derivation:
$$
\prooftree
  \prooftree
      {} \vdash_{\{j,l\}}
           \Client  \,{:}\,  {(
                            \Pi_i 
                                  \trity
                                    {{i{\mapsto}{-}}}
                                    {{i{\mapsto}{-}}}
                            \rightarrow
                            \Pi_i
                                  \trity
                                    {{i{\mapsto}{-}}}
                                    {\emp}
                        )} 
                         \rightarrow
                         \trity{j{\mapsto}{-}}
                               {\emp}
  \using 
    1 
  \justifies
  {} {\vdash_{\{j,l\}}}\,
       \Client  \,{:}\,  {\bigl((\Pi_i 
                        \trity
                           {{i{\mapsto}{-}}}
                           {{i{\mapsto}{-}}}
                      \rightarrow
                      \Pi_i
                         \trity
                           {{i{\mapsto}{-}}}
                           {\emp}) \otimes \inv(l)\bigr)}
                     \rightarrow
                     \trity{j{\mapsto}{-}{*}\inv(l)}
                           {\inv(l)} 
  \endprooftree
\justifies
  {} \vdash_{\{j,l\}} \Client\;\Mfree  \,{:}\,  
                     \trity{j{\mapsto}{-}*\inv(l)}
                           {\inv(l)} 
\endprooftree
$$
Here the step marked by $1$ is an instance of the third-order 
frame rule, and it applies the subsumption rule with the subtype
relation proved below:
$$
\begin{array}{l}
\qquad
  \bigl(\Pi_i \trity{{i{\mapsto}{-}}}{{i{\mapsto}{-}}}
         \rightarrow
         \Pi_i \trity{{i{\mapsto}{-}}}{\emp}\bigr) 
        \rightarrow \trity{j{\mapsto}{-}}{\emp}
\\
\preceq_{\{j,l\}}
\quad
(\because~  \theta \preceq_\Delta \theta \otimes P)
\\
\qquad
  \Bigl(
  \bigl(\Pi_i \trity{{i{\mapsto}{-}}}{{i{\mapsto}{-}}}
         \rightarrow
         \Pi_i \trity{{i{\mapsto}{-}}}{\emp}\bigr) 
        \rightarrow \trity{j{\mapsto}{-}}{\emp}
  \Bigr)\otimes \inv(l)
\\
\preceq_{\{j,l\}}
\quad
(\because~  
  (\theta\rightarrow \theta')\otimes P \simeq_\Delta 
  (\theta\otimes P \rightarrow \theta'\otimes P))
\\
\qquad
  \Bigl(\bigl(\Pi_i \trity{{i{\mapsto}{-}}}{{i{\mapsto}{-}}}
         \rightarrow
         \Pi_i \trity{{i{\mapsto}{-}}}{\emp}\bigr) \otimes \inv(l)\Bigr)
  \rightarrow
  \Bigl(\trity{j{\mapsto}{-}}{\emp} \otimes \inv(l)\Bigr)
\\
\preceq_{\{j,l\}}
\quad
(\because~  
  \trity{P}{Q}\otimes R \simeq_\Delta \trity{P*R}{Q*R})
\\
\qquad
  \Bigl(\bigl(\Pi_i \trity{{i{\mapsto}{-}}}{{i{\mapsto}{-}}}
         \rightarrow
         \Pi_i \trity{{i{\mapsto}{-}}}{\emp}\bigr) \otimes \inv(l)\Bigr)
  \rightarrow
  \trity{j{\mapsto}{-} * \inv(l)}{\emp *\inv(l)} 
\\
\preceq_{\{j,l\}}
\quad
(\because~  
 \forall \eta.\; \ff{P}_\eta = \ff{P*\emp}_\eta)
\\
\qquad
  \Bigl(\bigl(\Pi_i \trity{{i{\mapsto}{-}}}{{i{\mapsto}{-}}}
         \rightarrow
         \Pi_i \trity{{i{\mapsto}{-}}}{\emp}\bigr) \otimes \inv(l)\Bigr)
  \rightarrow
  \trity{j{\mapsto}{-}*\inv(l)}{\inv(l)}. 
\end{array}
$$

\section{Semantics}
\label{sec:semantics}

In this section we present our main contribution, 
the semantics that formalizes the underlying intuitions of
the separation-logic type system. In particular, we formalize
the following three intuitive properties of the type system:
\begin{enumerate}
\item The types in the separation-logic type system refine the
      conventional types. A separation-logic type 
      specifies a stronger property
      of a term, and restricts clients of such terms
      by asking them to only depend upon what can be known from
      the type. 
      For instance, the type $\trity{1\mapsto 3}{1\mapsto 0}$ of 
      a term $M$ indicates not just that $M$ is a command, but also that
      $M$ stores $0$ to cell $1$ if cell $1$ contains $3$ initially. 
      Moreover, this type forces clients to run $M$ only when
      cell $1$ contains $3$. 
\item The higher-order frame rules in the type system imply 
      that all programs behave locally.
\item The type system, however, does not change the computational 
      behavior of each program. 
\end{enumerate}
We formalize the first intuitive property by
means of partial equivalence relations.
Roughly, each type $\theta$ in our semantics determines
a partial equivalence relation (in short, per) 
over the meaning of the ``underlying type'' $\overline{\theta}$.
The domain of a per over a set $A$ is a subset of $A$;
this indicates that $\theta$ indeed 
specifies a stronger property than $\overline{\theta}$.
The other part of a per, namely the equivalence relation part,
explains that the type system restricts the clients, so that no
type-checked clients can tell apart two equivalent programs.
For instance, $\trity{1\mapsto 3}{1\mapsto 0}$ determines a per over the 
set of all commands. The domain of this per consists of commands satisfying
$\trity{1\mapsto 3}{1\mapsto 0}$, and the per equates two such commands
if they behave identically when cell $1$ contains $3$ initially.
The equivalence relation implies that type-checked clients
run a command of $\trity{1\mapsto 3}{1\mapsto 0}$ only when
cell $1$ contains $3$.

We justify the other two intuitive properties by proving 
technical lemmas about our semantics. For number 2, we prove the 
soundness of all the subtyping rules, including the generalized frame 
rule and the distribution rules. For number 3, we prove that our 
semantics has been obtained by extracting and then quotienting semantic elements
in the conventional semantics; yet, this extraction and quotienting
does not reduce the computational information of semantic elements.

In this section, we first define categories $\CC$ and $\cD$, corresponding
to the extraction and quotienting, respectively.  Next we give the
interpretation of types and terms. Finally, we connect our semantics with
the conventional semantics, and prove that our semantics is indeed obtained
by extracting and quotienting from the conventional semantics.

To make the paper accessible for a wider audience, we have decided to
present the categories $\CC$ and $\cD$ and the proofs of their properties
in a very concrete way --- it is possible to give equivalent, but more
abstract, descriptions of $\CC$ and $\cD$ and use known abstract results
>From category theory to prove some of their properties (e.g., cartesian
closure). 
For simplicity, we use the Hoare powerdomain to 
model the nondeterminism of commands in the semantics. Our results 
can be adapted to other alternatives, such as the Plotkin powerdomain
for countable nondeterminism, using the idea from the chapter 9.3.2 of
\cite{Yang-thesis}.

\subsection{Categories $\CC$ and $\cD$}\label{sec:categories-CD}
We construct $\CC$ and $\cD$ by modifying the category $\CPO$ of
pointed cpos and continuous functions. For $\CC$, we impose
a parameterized per
on each cpo, and extract only 
those morphisms in $\CPO$ that preserve such pers (at all instantiations).
The pers formalize that each type $\theta$ 
corresponds to a specification over the underlying type $\overline{\theta}$,
and the preservation of the pers guarantees that all
the morphisms in $\CC$ satisfy the corresponding specifications. 
The parameterization of each per gives an additional guarantee
that all morphisms in $\CC$ behave locally 
(in the sense of higher-order frame rules). 
The other category $\cD$ is a quotient of $\CC$. Intuitively,
the quotienting of $\CC$ reflects that our type system
also restricts the clients of a term; thus, more terms become
equivalent observationally.

We define the ``extracting'' category $\CC$ first.
Let $\spred$ be the set of {\em predicates}, i.e., subsets of $\heap$.
We recall the semantic version of separating connectives, 
$\semp$ and $*$, on $\spred$. For $p,q \in \spred$,
$$
\begin{array}{l}
  h \in \semp \iff h = \lambda n. \undefined, \\
  h \in p*q \iff \exists h_1h_2. \; h_1 \cdot h_2 = h \,\wedge\, 
                                    h_1 \in p \,\wedge\,
                                    h_2 \in q.
\end{array}
$$
The category $\CC$ is defined as follows:
\begin{itemize}
\item objects: $(A,R)$ where $A$ is a pointed cpo, and $R$ is a family
      of admissible pers\footnote{A per $R_0$ on $A$ is admissible iff
      $(\bot,\bot) \in R_0$ and $R_0$ is a sub-cpo of $A\times A$.} 
      indexed by predicates such that 
$$
      \forall p,q \in \spred.\, R(p) \subseteq R(p*q);
$$
\item morphisms: $f\colon (A,R) \rightarrow (B,S)$ is a continuous function
      from $A$ to $B$ such that 
$$
      \forall p \in \spred.\, f[R(p) \rightarrow S(p)]f,
$$
i.e., $f$ maps $R(p)$ related elements to $S(p)$ related elements.
\end{itemize}
Intuitively, an object $(A,R)$ denotes a specification parameterized
by invariant extension. The first component $A$ denotes
the underlying set from which we select ``correct'' elements.
$R(\semp)$ denotes the initial specification of this 
object where no invariant is added by the frame rule. The domain
$|R(\semp)|$ of per $R(\semp)$ indicates which elements satisfy the 
specification, and the equivalence relation on $|R(\semp)|$ expresses
how the specification is also used to limit the interaction of a client: 
the client can only do what the specification guarantees, so more
elements become equivalent observationally.
The per $R(p)$ at another predicate $p$ denotes an extended specification 
by the invariant $p$. 

We illustrate the intuition of $\CC$ with
a ``Hoare-triple'' object $[p,q]$ for $p,q \in \spred$. 
Let $\comm$ be the set of all functions $c$ from $\heap$ to 
$\cP(\heap \cup \{ \wrong \})$ that satisfy safety
monotonicity and the frame property:
\begin{itemize}
\item {\em Safety Monotonicity}: for all $h,h_0 \in \state$, 
      if $h\# h_0$ and $\wrong \not\in c(h)$,
      then $\wrong \not\in c(h\cdot h_0)$;
\item {\em Frame Property}: for all $h,h_0,h_1' \in \state$, if $h\# h_0$, 
      $\wrong \not\in c(h)$,
      and $h_1' \in c(h \cdot h_0)$, then there exists $h'$ such that
      $h_1' = h' \cdot h_0$ and $h' \in c(h)$.
\end{itemize}
The above two properties are from the work on separation logic,
and they form a sufficient and necessary condition that commands 
satisfy the (first-order) frame rule \cite{YO02}. Note that
the safety monotonicity and frame property are
equivalent to the following condition:\footnote{The
inclusion is one way only. For a counterexample,
consider two disjoint heaps $h{=}[1\bind 0]$
and $h_0 {=} [2\bind 0]$ and the command 
$$
\;\vdash\;
\mylet{j{=}\new}{\bigl(\free(j);(\ifz\; (j{-}2)\; ([1]:=5)\;([1]:=6))\bigr)}
: 
\trity{1\mapsto-}{1\mapsto-}.
$$
When this command is run in $h$, it nondeterministically
assigns $5$ or $6$ to location $1$, but when it is run in a bigger
heap $h\cdot h_0$, the command always assigns $6$ to the same location.
}
$$
\mbox{if $h\# h_0$ and $\wrong$ isn't in $c(h)$, then 
$c(h \cdot h_0) \subseteq \{ h' \cdot h_0 \mid h' \in c(h) \mbox{ and } h'\#h_0 \}$.}
$$
The set $\comm$ is the first component of the Hoare-triple object $[p,q]$,  
where the order on $\comm$ is given by:
$$
    c \sqsubseteq c' 
    \iff 
    \forall h.\, c(h) \subseteq c'(h). 
$$

The real meaning of $[p,q]$ is given by the second component $R$.
For each predicate $p_0$, 
the domain of $R(p_0)$ consists of all ``commands'' in $\comm$ that 
satisfy $\trity{p*p_0}{q*p_0}$:
$$
   c \in |R(p_0)|  \iff \forall h \in p*p_0.\, c(h) \subseteq q*p_0.
$$
The equivalence relation $R(p_0)$ relates $c$ and $c'$ in $|R(p_0)|$
iff $c$ and $c'$ behave the same for the inputs in $p*p_0*\strue$:
$$
\begin{array}{rcl}
    \strue       & = &  \{h \mid h \in \heap\}
\\
    c[R(p_0)]c' & \iff &
    \forall h \in p*p_0*\strue.\, c(h) = c'(h).
\end{array}
$$
This equivalence relation means that the type
system allows a client to execute $c$ or $c'$ in $h$
only when $h$ satisfies $p*p_0*p'$ for some $p'$, which is added
by the frame rule. We remark that the $*$ operator
in the definition of $|R(p_0)|$ is allowed to
partition the heap differently before and after the execution of $c$. 
For instance, when 
$$
    p = \{[1\bind 1]\},\;
    q = \{[2\bind 0]\},\;
    \mbox{and}\; 
    p_0 = \{[2\bind 0, 3\bind 0],\; [1\bind 0,3\bind 1]\},
$$
the initial heap $h$ in the definition is split into
cell $1$ for $p$ and cells $2,3$ for $p_0$, but the final heap 
is split into cell $2$ for $q$ and cells $1,3$ for $p_0$.

The category $\CC$ is cartesian closed, has all small products,
and contains the least fixpoint operator.
The terminal object is $(\{\bot\},\botcomp)$ where $\botcomp(p)$ 
is $\{(\bot,\bot)\}$ 
for all $p$, and the small products are given pointwise; for instance,
$(A,R)\times (B,S)$ is $(A\times B, \{R(p)\times S(p)\}_p)$. The exponential 
of $(A,R)$ and $(B,S)$ is subtle, and its per component involves the 
quantification over all predicates. The 
cpo component of the exponential $(A,R) \Rightarrow (B,S)$ 
is the continuous function space
$A\Rightarrow B$, and the per component of
$(A,R) \Rightarrow (B,S)$, denoted $R\Rightarrow S$,
is defined as follows:
$$
\begin{array}{rcl}
  f \in |(R\Rightarrow S)(p)| & \iff & 
  \forall q \in \spred.\, f[R(p*q)\rightarrow S(p*q)]f,
\\
  f[(R\Rightarrow S)(p)]g & \iff & f,g\in |(R\Rightarrow S)(p)|
  \;\;\wedge\;\;
  \forall q \in \spred.\, f[R(p*q)\rightarrow S(p*q)]g.
\\
\end{array}
$$
Note that the right hand sides of the above equivalences
quantify over all $*$-extension $p*q$ of $p$.
This quantification ensures that $R\Rightarrow S$ satisfies
the requirement 
$$
  \forall p,p' \in \spred.\;
  (R\Rightarrow S)(p) \subseteq (R\Rightarrow S)(p*p')
$$
in the category $\CC$.
\begin{lem}
$\CC$ is cartesian closed, and has all small products.
\end{lem}
\proof
First, we prove that 
for every (small) family $\{(A_i,R_i)\}_{i \in I}$ of objects in $\CC$,
its product is $(\Pi_{i\in I}A_i,\Pi_{i\in I}R_i)$
and the $i$-th projection $\pi_i$ is $\lambda x. x(i)$. 
Here we write $(\Pi_{i\in \emptyset}A_i,\Pi_{i\in \emptyset}R_i)$
for $(\{\bot\},\botcomp)$. It is straightforward to show that 
$(\Pi_{i\in I}A_i,\Pi_{i\in I}R_i)$ is
an object in $\CC$ and $\pi_i$ is a morphism in $\CC$.
So, we focus on proving the usual universality requirement for
the product. Consider
an object $(B,S)$ and a family ${f_i:(B,S)\rightarrow (A_i,R_i)}_{i \in I}$
of morphisms in $\CC$. We need to prove that there exists a unique
morphism $k$ from $(B,S)$ to $(\Pi_{i\in I}A_i,\Pi_{i\in I}R_i)$,
such that 
$$
    \forall i \in I.\; f_i = \pi_i \circ k.
$$
The above formula is equivalent to saying that $k$ is 
$g = \lambda b. \lambda i. f_i(b)$. In particular, 
when $I = \emptyset$, $k$ has
to be the unique function $g' = \lambda b. \bot$. Note that
these characterizations give the uniqueness of $k$.
We prove the existence of $k$, by showing that $g$ and $g'$ are
morphisms in $\CC$.  The continuity of $g$ and $g'$ is well-known.
The relation preservation of $g'$ also easily follows, since
$\botcomp$ is a family of complete relations. For
the relation preservation of $g$,
we use the fact that $f_i$'s are
the morphisms in $\CC$.
Pick an
arbitrary predicate $p$, and choose $b,b'$ from $B$ such that
$b[S(p)]b'$. Then,
$$
\begin{array}{r@{\;}c@{\;}l}
  \forall i \in I.\; f_i(b)[R_i(p)]f_i(b')
&
\iff
&
  \forall i \in I.\; g(b)(i)\bigl[R_i(p)\bigr]g(b')(i)
\hfill
\qquad
(
\because~ 
\mbox{the definition of $g$}
)
\\
&
\iff
&
  g(b)\bigl[(\Pi_{i\in I}R_i)(p)\bigr]g(b')
\hfill
\qquad
(
\because~ 
\mbox{the definition of $\Pi_{i\in I}R_i$}
).
\end{array}
$$

Next, we prove that $(A \Rightarrow B,R\Rightarrow S)$ is an
exponential of $(A,R)$ and $(B,S)$, with the evaluation morphism
$\ev = \lambda (f,x).f(x)$. 
It is straightforward to prove
that $\ev$ is a morphism in $\CC$
and $(A \Rightarrow B,R\Rightarrow S)$ is an object in $\CC$.
So, we focus on the universality
requirement for the exponentials. Consider a morphism 
$f\colon (C,T)\times (A,R) \rightarrow (B,S)$ in $\CC$. We need to
show that there exists a unique morphism 
$\curry(f):(C,T) \rightarrow (A\Rightarrow B,R\Rightarrow S)$ such that 
$$
   \forall (c,a) \in C\times A.\;
   f(c,a) = \ev(\curry(f)(c), a).
$$
Since $\ev(\curry(f)(c), a) = \curry(f)(c)(a)$,
the above is equivalent to
$\curry(f) = \lambda c.\lambda a.f(c,a)$.
Note that this characterizes $\curry(f)$ completely, so
it gives the uniqueness of $\curry(f)$. It remains to prove that
$\curry(f)$ is a morphism in $\CC$. It is well-known that $\curry(f)$
is a continuous function from $C$ to $A\Rightarrow B$. Thus,
we only prove the relation preservation of $\curry(f)$,
using the fact that $f[T(p)\times R(p) \rightarrow S(p)]f$ for all $p$.  Pick arbitrary predicate $p$ and $c,c'$ in $C$
such that $c[T(p)]c'$. Then, for all predicates $q$,
we have that $c[T(p*q)]c'$, because $T(p) \subseteq T(p*q)$. Thus,
$$
\begin{array}{l}
\qquad
\forall q.\forall a,a' \in A.\; a[R(p*q)]a' \implies f(c,a)[S(p*q)]f(c',a')
\\
\iff
\quad
(
\because~ 
\mbox{the definition of $R\Rightarrow S$}
)
\\
\qquad
(\lambda a.f(c,a)) \bigl[(R \Rightarrow S)(p)\bigr] (\lambda a'.f(c',a'))
\\
\iff
\quad
(
\because~ 
\mbox{the definition of $\curry(f)$}
)
\\
\qquad
\curry(f)(c) \bigl[(R \Rightarrow S)(p)\bigr] \curry(f)(c').
\end{array}
$$
\qed

\begin{lem}
For every object $(A,R)$ in $\CC$, the least fixpoint
operator $\lfix_A \colon [A\Rightarrow A] \rightarrow A$ on $A$
is a morphism in $\CC$.
\end{lem}
\proof
Pick arbitrary predicate $p$, and continuous functions 
$f,g$ of type $A \rightarrow A$, such that
$f[(R \Rightarrow R)(p)]g$; equivalently, 
$f[R(p*q)\rightarrow R(p*q)]g$ for all $q$. 
We need to show that $\lfix(f)[R(p)]\lfix(g)$. Note that since $R$ is 
admissible, it is sufficient to prove that 
$f^k(\bot)[R(p)]g^k(\bot)$ for all $k \geq 0$. 
This sufficient condition holds because
$f[R(p)\Rightarrow R(p)]g$ and $\bot[R(p)]\bot$.
\qed

Another important feature of $\CC$ is that it validates
higher-order frame rules. Let $\cPr$ be the
preorder $(\spred, \sqsubseteq)$ with
$\sqsubseteq$ defined by predicate extension:
$$
        p \sqsubseteq r \iff \exists q. p*q = r.
$$
Category $\CC$ has an ``invariant-extension'' functor $\inv$ from 
$\CC \times \cPr$ to $\CC$ defined by:
$$
   \inv((A,R),p) = (A,R(p*-)) \;\mbox{ and }\; \inv(f,p\sqsubseteq q) = f.
$$
Functor $\inv$ corresponds to the type constructor $\otimes$ in our
language; given a ``type'' $(A,R)$ and a predicate $p$, $\inv$ extends 
$(A,R)$ by adding the invariant $p$. For instance, when a triple 
object $[p',q']$ is extended with $p$, it becomes $[p'*p,q'*p]$. 

Functor $\inv$ validates the subtyping rules that express higher-order
frame rules: the generalized frame rule $\theta \preceq_\Delta \theta\otimes P$
and the rules for distributing $\otimes$ over each type constructor.
We first show that the functoriality of $\inv$
gives the soundness of the generalized frame rule.
Note that $\semp \sqsubseteq p$ for all predicates $p$, 
and that $\inv(-,\semp)$ is the identity functor 
on $\CC$. Thus, for each $(A,R)$, the functoriality of $\inv$ gives a 
morphism from $(A,R)$ to $\inv((A,R),p)$. This morphism gives the soundness
of the subtyping rule $\theta \preceq_\Delta \theta\otimes P$. 

The soundness of the other distribution rules follows from the fact
that for all $p$, $\inv(-,p)$ preserves most of the structure of $\CC$.
For instance, $\inv(-,p)$ preserves 
the exponential of $\CC$, because for all objects $(A,R)$ and $(B,S)$
and all predicates $q$,
we have that
$$
\begin{array}{rcl}
   f[(R(p*-)\Rightarrow S(p*-))(q)]g 
&
\iff
&
   \forall q'. f[R(p*(q*q'))\rightarrow S(p*(q*q'))]g 
\\
&
\iff
&
   \forall q'. f[R((p*q)*q')\rightarrow S((p*q)*q')]g 
\\
&
\iff
&
   f[(R\Rightarrow S)(p*q)]g.
\end{array}
$$
\begin{lem}\label{lemma:inv-preservation}
For each predicate $p$, $\inv(-,p)$ preserves the cartesian closed structure
and all the small products of $\CC$ on the nose.
\end{lem}
\proof
It is sufficient to prove that $\inv(-,p)$ preserves 
exponential objects, small product objects,
evaluation morphisms, and projection morphisms. 
First, we prove the preservation of the small product objects
and projection morphisms.
Consider a family $\{(A_i,R_i)\}_{i\in I}$ of objects
in $\CC$. The following shows that the product 
$\Pi_{i \in I}(A_i,R_i)$ of this family is preserved by 
$\inv(-,p)$:
$$
\begin{array}{r@{\;}c@{\;}l}
\inv\bigl(\Pi_{i \in I}(A_i,\, R_i),\; p\bigr) 
&
=
&
\inv\bigl(\bigl(\Pi_{i \in I}A_i,\, \Pi_{i\in I}R_i\bigr),\; p\bigr) 
\quad
\hfill
(\because~ 
\mbox{the definition of products in $\CC$}
)
\\
&
=
&
\bigl(\Pi_{i \in I}A_i,\, (\Pi_{i\in I}R_i)(p*-)\bigr) 
\quad
\hfill
(\because~ 
\mbox{the definition of $\inv$}
)
\\
&
=
&
\bigl(\Pi_{i \in I}A_i,\, \Pi_{i\in I}(R_i(p*-))\bigr) 
\\
&
=
&
\Pi_{i \in I}\bigl(A_i,\, R_i(p*-)\bigr)
\\
&
=
&
\Pi_{i \in I}\bigl(\inv((A_i,\,R_i),\;p)\bigr)
\quad
\hfill
(\because~ 
\mbox{the definition of $\inv$}
).
\end{array}
$$
Since $\inv(f,p) = f$, functor $\inv$ preserves the $i$-th
projection from $\Pi_{i \in I}(A_i,R_i)$.

Next, we show that $\inv(-,p)$ preserves the exponential objects and
evaluation morphisms in $\CC$.
Let $(A,R)$ and $(B,S)$ be objects in $\CC$. By what we have shown
before this lemma, we have that
$$
   (R \Rightarrow S)(p*-) \;=\; R(p*-) \Rightarrow S(p*-).
$$
>From this follows the preservation of exponential objects:
$$
\begin{array}{@{}r@{\;}c@{\;}l@{}}
\inv\bigl((A,R) \Rightarrow (B,S),\; p\bigr)
&
= 
& 
\inv\bigl((A\Rightarrow B,\,R\Rightarrow S),\; p\bigr)
\;
\hfill
(\because 
\mbox{Def. of exponentials in $\CC$}
)
\\
&
= 
& 
\bigl(A\Rightarrow B,\; (R\Rightarrow S)(p*-)\bigr)
\;
\hfill
(\because 
\mbox{Def. of $\inv$}
)
\\
& 
= 
& 
\bigl(A\Rightarrow B,\; (R(p*-) \Rightarrow S(p*-))\bigr)
\\
& 
= 
& 
\bigl(A,R(p*-)\bigr) \Rightarrow 
\bigl(B,S(p*-)\bigr)
\;
\hfill
(\because 
\mbox{Def. of exponentials in $\CC$}
)
\\
& 
= 
&
\inv\bigl((A,R),p\bigr) \Rightarrow 
\inv\bigl((B,S),p\bigr)
\;
\hfill
(\because 
\mbox{Def. of $\inv$}
).
\end{array}
$$
Functor $\inv(-,p)$ preserves the 
the evaluation morphism 
$\ev$ for $(A,R)\Rightarrow (B,S)$,
 because $\inv(-,p)$ preserves the products and exponentials
and $\inv(f,p)$ only changes the type of $f$,
not modifying its ''meaning'' 
(i.e., $\inv(f,p) = f$).
\qed

\begin{lem}\label{lemma:inv-preservation2}
For all predicates $p$ and $q$, $\inv(-,p) \circ \inv(-,q) = \inv(-,p*q)$.
\end{lem}
\proof
Both $\inv(-,p) \circ \inv(-,q)$ and $\inv(-,p*q)$ map
a morphism $f$ to the same $f$ with perhaps different domain and 
codomain. Thus, if they act the same on 
the objects in $\CC$, they must act the same on the morphisms.
In fact, they do act the same on the objects; for each $(A,R)$ in $\CC$,
$$
\begin{array}{rcl}
(\inv(-,p) \circ \inv(-,q))(A,R) 
& = & 
(A,R(p*(q*-))) 
\hfill
\qquad
(\because~  \mbox{the definition of $\inv$})
\\
& = &
(A,R((p*q)*-)) 
\hfill
\qquad
(\because~  \mbox{$*$ is associative})
\\
& = &
\inv(-,p*q)(A,R)
\hfill
\qquad
(\because~  \mbox{the definition of $\inv$}).
\end{array}
$$
\qed

For now, the final remark on $\CC$ is that the triple-object generator $[-,-]$ 
can be made into a functor, whose morphism action validates
the subtyping rule for Consequence. Let $\cP$ be the set of predicates
ordered by the subset inclusion $\subseteq$. Generator $[-,-]$
can be extended to a functor $\tri$ from $\cP^\op\times \cP$ to $\CC$:
$$
   \tri(p,q) = [p,q] \;\mbox{ and }\;
   \tri(p'\subseteq p,q\subseteq q')(c) = c.
$$
Note that $\tri$ is contravariant in the first argument and
covariant on the second argument. This mixed variance reflects
that the pre-condition of a triple can be strengthened, and the
post-condition can be weakened; thus, it validates
the subtyping rule for Consequence. We also note that the subtyping
rule that moves an invariant assertion into the pre- and post-conditions
is sound.
\begin{lem}\label{lemma:inv-preservation3}
For each predicate $p$, let $-*p\colon \cP \rightarrow \cP$
be a functor that maps a predicate $q$ to $q*p$. Then, 
$$
        \inv(-,p)\circ \tri = \tri(-*p,-*p).
$$
\end{lem}
\proof
Both $\inv(-,p)\circ\tri$ and $\tri(-*p,-*p)$ map
the morphisms in $\cP^\op \times\cP$ to inclusions
between pointed cpos. Thus, it is sufficient to prove that
$\inv(-,p)\circ\tri$ and $\tri(-*p,-*p)$ act the same
on objects. Pick an arbitrary object
$(p',q')$ in $\cP^\op \times\cP$. Then, by the definition of
$\inv$ and $\tri$, there exist families $R,S$ of pers such that
$$
   (\comm,R) =
   (\inv(-,p)\circ \tri)(p',q')
   \,\;\mbox{ and }\;\;
   (\comm,S) = 
   (\tri(-*p,-*p))(p',q').
$$
Thus, to prove $(\inv(-,p)\circ \tri)(p',q') = (\tri(-*p,-*p))(p',q')$,
we only need to show  $R=S$. For each predicate $p_0$, 
the domains of $R(p_0)$  and $S(p_0)$ are the same, because
$$
\begin{array}{r@{\;}c@{\;}l}
  c \in |R(p_0)| 
&
  \iff
&
  \forall h \in p'*(p*p_0).\; c(h) \subseteq q'*(p*p_0)
\hfill
\;\;
(\because~  \mbox{Def. of $\inv(\tri(p',q'),p)$})
\\
&  \iff
&
  \forall h \in (p'*p)*p_0.\; c(h) \subseteq (q'*p)*p_0
\hfill
\;\;
(\because~  \mbox{$*$ is associative})
\\
& \iff
&
  c \in |S(p_0)| 
\hfill
\;\;
(\because~  \mbox{Def. of $\tri(p'*p,q'*p)$}).
\end{array}
$$
And, $R(p_0)$ and $S(p_0)$ specify the same relation on their domains,
because
$$
\begin{array}{r@{\;}c@{\;}l}
  c[R(p_0)]c'
&
  \iff
&
  \forall h \in p'*(p*p_0)*\strue.\; c(h) = c'(h)
\hfill\;\;
(\because~ \mbox{Def. of $\inv(\tri(p',q'),p)$})
\\
&
  \iff
&
  \forall h \in (p'*p)*p_0*\strue.\; c(h) = c'(h)
\hfill\;\;
(\because~ \mbox{$*$ is associative})
\\
&
  \iff
&
  c[S(p_0)]c'
\hfill\;\;
(\because~ \mbox{Def. of $\tri(p'*p,q'*p)$}).
\end{array}
$$
\qed

The category $\cD$ is obtained from $\CC$ by equating
 morphisms according to an equivalence relation $\sim$. 
Morphisms $f$ and $g$ in $\CC[(A,R),(B,S)]$ are related by $\sim$ iff
$$
   \forall p \in \spred.\,f[R(p)\rightarrow S(p)]g.
$$
$\sim$ is an equivalence relation; it is reflexive,
because every morphism in 
$\CC[(A,R),(B,S)]$ should map $R(p)$-related elements to $S(p)$-related
elements, for all $p$; it is symmetric and transitive
because, for all $p$, $R(p)$ and $S(p)$ are symmetric and transitive.
The interesting property of $\sim$ is that it is preserved by 
all the structure of $\CC$:
\begin{lem}[Preservation]\label{lemma:preservation-quotient}
The relation $\sim$ is preserved by the following operators in $\CC$:
\begin{itemize}
\item the functor $\inv(-,p\sqsubseteq q)$ on $\CC$,
      for all predicates $p,q$ such that $p\sqsubseteq q$;
\item the composition of morphisms;
\item the currying of morphisms; and
\item the pairing into all the small products.
\end{itemize}
\end{lem}
\proof
First, we prove the preservation by $\inv$. Let $p$ and $q$
be predicates such that $p\sqsubseteq q$. Pick arbitrary 
two morphisms $f,g \colon (A,R)\rightarrow (B,S)$ in $\CC$ such that 
$f\sim g$. We will show that 
$\inv(f,p\sqsubseteq q) \sim \inv(g,p\sqsubseteq q)$. 
Morphism $\inv(f,p\sqsubseteq q)$ and $\inv(g,p\sqsubseteq q)$
both have the type $(A,R(p*-)) \rightarrow (B,S(q*-))$. Thus,
proving $\inv(f,p\sqsubseteq q) \sim \inv(g,p\sqsubseteq q)$
amounts to showing the formula:
$$
\forall r \in \spred.\;
f[R(p*r) \rightarrow S(q*r)]g.
$$
The formula holds, because $f[R(p*r) \rightarrow S(p*r)]g$ 
and $S(p*r) \subseteq S(q*r)$ for all $r$.

Second, we prove the preservation by the composition of morphisms.
Consider morphisms $f,f'\colon (A,R) \rightarrow (B,S)$ and
$g,g'\colon (B,S) \rightarrow (C,T)$ such that $f\sim f'$ and $g\sim g'$.
Then, for all predicates $p$ and all $a,a'\in A$ such that $a[R(p)]a'$,
we have that $f(a)[S(p)]f'(a')$, so $g(f(a))[T(p)]g'(f'(a'))$. This proves
that $(g\circ f)\sim (g'\circ f')$.

Third, we show the preservation by the currying operator. Consider
morphisms $f,f'$ from $(C,T)\times (A,R)$ to $(B,S)$, such that
$f\sim f'$. Pick an arbitrary predicate $p$, and choose $T(p)$-related
$c,c'$ from $C$. Then, for all predicates $q$, we have that
$c[T(p*q)]c'$, because $T(p) \subseteq T(p*q)$. Thus, 
$$
\begin{array}{l}
\qquad
  \forall q.\forall a,a' \in A.\;
  a[R(p*q)]a' \Rightarrow f(c,a)[S(p*q)]f'(c',a')
\\
\iff
\quad
(
\because~ 
\mbox{the definition of $R\Rightarrow S$}
)
\\
\qquad
  (\lambda a. f(c,a))\bigl[(R \Rightarrow S)(p)\bigr](\lambda a'. f'(c',a'))
\\
\iff
\quad
(
\because~ 
\mbox{the definition of $\curry(f)$}
)
\\
\qquad
  \curry(f)(c)\bigl[(R \Rightarrow S)(p)\bigr]\curry(f')(c')
\end{array}
$$
What we have just proved shows that $\curry(f)\sim\curry(f')$.

Finally, we prove the preservation by the pairing into the small products. 
Consider a family $\{(A_i,R_i)\}_{i \in I}$ of objects in $\CC$. Pick
two families of morphisms in $\CC$,  $\{f_i\}_{i\in I}$
and $\{f'_i\}_{i\in I}$,  such that
$$
\forall i \in I.\;\;\;
f_i\colon (B,S) \rightarrow (A_i,R_i),\;\;\;
f'_i\colon (B,S) \rightarrow (A_i,R_i),
\;\;\;\mbox{and}\;\;\;
f_i\sim f'_i.
$$ 
We need to show the following equivalence:
$$
   \Bigl(\lambda b\colon B.\lambda i\colon I. f_i(b)\Bigr)
   \sim
   \Bigl(\lambda b'\colon B.\lambda i\colon I. f'_i(b')\Bigr).
$$
For all predicates $p$ and all $b,b'$ in $B$ such that $b[S(p)]b'$,
we have that
$$
   \forall i \in I.\; f_i(b)[R_i(p)]f'_i(b').
$$
Thus, 
$$
   (\lambda i\colon I.\, f_i(b))[(\Pi_{i\in I}R_i)(p)]
   (\lambda i\colon I.\, f'_i(b')).
$$
This relationship gives the required equivalence.\qed\nobreak
\nobreak Lemma~\ref{lemma:preservation-quotient} 
ensures that taking a quotient of morphisms in $\CC$ gives
a well-defined category, which we call $\cD$. 
Category $\cD$ inherits
all the interesting structure of $\CC$ by
Lemma~\ref{lemma:preservation-quotient};\break 
it is cartesian closed, has all small products, 
 and has a functor 
$\inv'\colon \cD\times\cPr\rightarrow \cD$ that pre-\break serves
the CCC structure and the small products of $\cD$. Let $E$ be
the ``quotienting''functor from $\CC$ to $\cD$,
and $\tri'\colon \cP^\op\times \cP \rightarrow \cD$ the composition
of $E$ with $\tri$. We summarize the main property of $\cD$ in the following
two lemmas:

\break\begin{lem}
\label{lem:cD}
The category $\cD$ is a CCC with all small products, and has
two functors $\inv'\colon \cD \times \cPr \rightarrow \cD$ and
$\tri' \colon \cP^\op\times \cP \rightarrow \cD$ such that
\begin{enumerate}
\item $\inv'(-,p)$ preserves all the CCC structure and the small products of 
$\cD$;
\item\label{lem:cD:i2}
  $\inv'(-,p) \circ \inv'(-,q) = \inv'(-,p*q)$; and
\item \label{lem:cD:i3}
$\inv'(-,p) \circ \tri' = \tri'(-*p,-*p)$.
\end{enumerate}
\end{lem}
\proof
First, we prove that $\cD$ has all the small products.
Let $\{(A_i,R_i)\}_{i \in I}$ be a small family of objects in $\cD$.
We show that the product of this family is $(\Pi_{i\in I}A_i,\Pi_{i\in I}R_i)$
and the $i$-th projection is $[\pi_i]$, where $[f]$ means the equivalence
class of the morphism $f$. Consider an arbitrary family 
$\{[f_i]\colon (B,S) \rightarrow (A_i,R_i)\}_{i \in I}$ of morphisms
in $\cD$. This family induces some family $\{f_i \}_{i \in I}$ in $\CC$. 
Since $(\Pi_{i\in I}A_i,\Pi_{i\in I}R_i)$ is the product in $\CC$,
there exists a morphism 
$\langle f_i \rangle_{i\in I}
 \colon (B,S) \rightarrow (\Pi_{i\in I}A_i,\Pi_{i\in I}R_i)$
such that $\pi_i \circ \langle f_i\rangle_{i\in I} = f_i$ for all $i\in I$. The equivalence class
$[\langle f_i\rangle_{i\in I}]$ of this morphism is the required 
unique morphism in $\cD$.
It makes the required diagrams for the products commute, because 
$$
  \forall i \in I.\;
  [\pi_i] \circ [\langle f_i \rangle_{i\in I}] = 
  [\pi_i \circ \langle f_i \rangle_{i \in I}] =
  [f_i].
$$
For the uniqueness, suppose that $[k]$ is another morphism in $\cD$ 
that makes the diagram commutes. Then, $[k]$ must be equal
to $[k] = [\langle f_i \rangle_{i\in I}]$, as shown below:
$$
\begin{array}{@{}r@{\;}c@{\;}l@{}}
  (\forall i \in I.\; [\pi_i] \circ [k] = [f_i])
& 
\iff
&
  (\forall i \in I.\; [\pi_i \circ k] = [f_i]) 
\hfill
\;\;
(\because 
 \mbox{$\sim$ is preserved by $\circ$}
)
\\
& 
\implies
&
  [\langle\pi_i \circ k\rangle_{i\in I}] = [\langle f_i \rangle_{i\in I}] 
\hfill
\;\;
(\because 
 \mbox{$\sim$ is preserved by the pairing}
)
\\
& 
\iff
&
  [k] = [\langle f_i \rangle_{i\in I}].
\end{array}
$$

Second, we show that $\cD$ has the exponentials. Let $((A,R),(B,S))$
be a pair of objects in $\cD$. We prove 
that $(A\Rightarrow B,R\Rightarrow S)$ 
is an exponential of this pair, and the evaluation morphism is the
equivalence class $[\ev]$. Consider a morphism 
$[f]\colon (C,T)\times(A,R) \rightarrow (B,S)$ in $\cD$. We need
to prove that the universality requirement holds for $[f]$: 
there exists a unique morphism 
$[g]\colon (C,T) \rightarrow (A\Rightarrow B,R\Rightarrow S)$ in $\cD$ 
such that
$$
   [f] = [\ev] \circ \langle [g] \circ [\pi_0], [\pi_1]\rangle.
$$
The equation in the requirement implies that $[g]$ should be
equal to $[\curry(f)]$:
$$
\begin{array}{@{}r@{\;}c@{\;}l@{}}
   [f] = [\ev] \circ \langle [g] \circ [\pi_0], [\pi_1]\rangle
&
\implies
&
[f] = [\ev] \circ \langle [g \circ \pi_0], [\pi_1]\rangle
\hfill\;\;
(\because
\mbox{the composition preserves $\sim$}
)
\\
&
\implies
&
[f] = [\ev] \circ [\langle g \circ \pi_0, \pi_1\rangle]
\hfill\;\;
(\because 
\mbox{the pairing preserves $\sim$}
)
\\
&
\implies
&
   [f] = [\ev \circ \langle g \circ \pi_0, \pi_1\rangle]
\hfill
\;\;
(\because
\mbox{the composition preserves $\sim$}
)
\\
&
\implies
&
   [\curry(f)] = [\curry(\ev \circ \langle g \circ \pi_0, \pi_1\rangle)]
\hfill\;\;
(\because 
\mbox{$\curry$ preserves $\sim$}
)
\\
&
\implies
&
   [\curry(f)] = [g].
\end{array}
$$
Thus, $\cD$ has at most one morphism $[g]$ that satisfies
the universality requirement. We now show that $[\curry(f)]$ satisfies 
the requirement. By the definition of $\curry$, 
we have that
$$
    f = \ev \circ \langle \curry(f) \circ \pi_0, \pi_1\rangle.
$$
This equation implies that $[\curry(f)]$ makes the required
diagram commute:
$$
\begin{array}{@{}r@{\,}c@{\,}l@{}}
    f = \ev \circ \langle \curry(f) \circ \pi_0, \pi_1\rangle
&
\implies
&
    [f] = [\ev \circ \langle \curry(f) \circ \pi_0, \pi_1\rangle]
\\
&
\implies
&
    [f] = [\ev] \circ [\langle \curry(f) \circ \pi_0, \pi_1\rangle]
\hfill\;
(\because
\mbox{$\circ$ preserves $\sim$}
)
\\
&
\implies
&
    [f] = [\ev] \circ \langle [\curry(f) \circ \pi_0], [\pi_1]\rangle
\hfill\;
(\because 
\mbox{pairing preserves $\sim$}
)
\\
&
\implies
&
    [f] = [\ev] \circ \langle [\curry(f)] \circ [\pi_0], [\pi_1]\rangle
\hfill\;
(\because
\mbox{$\circ$ preserves $\sim$}
).
\end{array}
$$

Finally, we prove the three properties of $\inv'$. Note that 
the categories $\CC$ and $\cD$ have the same collection of 
objects, and they have the same exponentials and same small products,
as far as the objects are concerned. Moreover,
for objects, the functors $\inv'(-,p)$ and $\inv(-,p)$
are identical. Thus, $\inv'(-,p)\colon \cD \rightarrow \cD$ preserves
the exponential objects and small product objects in $\cD$ if and only if
 $\inv(-,p)$ preserves those in $\CC$; the right hand side of this equivalence 
holds
by Lemma~\ref{lemma:inv-preservation}. The functor $\inv'(-,p)$
also preserves $[\ev]$ and $[\pi_i]$, because
$\inv'([f],p) = [\inv(f,p)] = [f]$. So,
$\inv'(-,p)$ preserves the CCC structure
and the small products.

For the second property of $\inv'$, we note that 
the equation in the property holds for the objects,
because for all predicates $r$, functors $\inv'(-,r)$ and $\inv(-,r)$ 
behave the same on the objects, and 
$\inv(-,p) \circ \inv(-,q) = \inv(-,p*q)$. The equation also holds
for the morphisms, because $\inv'([f],r) = [f]$ for all $f,r$.

For the third property of $\inv'$, we recall that $\tri' = E \circ \tri$.
Thus, it is sufficient to show that
$$
    \inv'(-,p) \circ E \circ \tri = E \circ \tri(-*p,-*p).
$$
The equation holds for the objects; 
$E$ is the identity on the objects, $\inv$ and $\inv'$ are
the same for objects, and $\inv(-,p)\circ\tri = \tri(-*p,-*p)$.
For the morphisms, the equation also holds, because
both sides of the equation map each morphism in $\cP^\op\times\cP$ 
to the equivalence class
of an inclusion.
\qed
\begin{lem}\label{lemma:preservation-E}
The functor $E$ from $\CC$ to $\cD$ is full, 
preserves the CCC structure as well as small products,
and makes the following diagrams commute:
$$
\xymatrix{
\CC \times \cPr   \ar[r]^-\inv  \ar[d]_{E\times\Id} & \CC \ar[d]^E \\
\cD \times \cPr   \ar[r]^-{\inv'} & \cD
}
\xymatrix{
\cP^\op \times \cP   \ar[r]^-\tri \ar[d]_{\Id} & \CC \ar[d]^E \\
\cP^\op \times \cP   \ar[r]^-{\tri'} & \cD
}
$$
\end{lem}
\proof
The categories $\CC$ and $\cD$ have the same collection of objects,
and their CCC structure and small products are identical, as far
as the objects are concerned. Since $E$ is the identity on objects,
it preserves the exponential objects and small product objects.
Moreover, $E$ preserves the evaluation and projection morphisms,
because the evaluation and projection morphisms in $\cD$ are just
the equivalence classes of the corresponding morphisms in $\CC$,
and $E$ maps $f$ to its equivalence class $[f]$. Thus,
functor $E$ preserves the CCC structure and the small products of $\CC$.

The commutative diagram for $\inv$ holds for the objects, because
$\inv$ and $\inv'$ behave the same for the objects and $E$ is the
identity on the objects. To show that the diagram also holds for 
the morphisms, we pick an arbitrary morphism $(f,p\sqsubseteq q)$ in 
$\CC\times \cPr$. Then,
$$
   (\inv' \circ (E\times \Id))(f,p\sqsubseteq q) 
   \;\;=\;\; 
   \inv'([f],p\sqsubseteq q) 
   \;\;=\;\;
   [f] 
   \;\;=\;\; 
   (E \circ \inv)(f,p\sqsubseteq q).
$$

Finally, the commutative diagram for $\tri'$ is the definition of $\tri'$,
so it must hold.
\qed

\subsection{Interpretation of the Language}\label{sec:interpretation-pl}
We interpret the language in two steps. First, we define the 
semantics $\cff{-}$ in the family fibration $\Fam(\CC) \rightarrow \Set$.
Each base set in the fibration
models all the possible environments for a fixed shape of the stack 
(i.e., a fixed set of integer variables $\Delta$).  For instance, the object
$\{(A,R)_\eta\}_{\eta \in \ff{\Delta}}$ assumes that all the available
integer variables are in $\Delta$, and it specifies a type dependent
on the values of such variables, given by $\eta$.  The types and
terms of our language are interpreted 
using the categorical structure of the fibration.
Next, we quotient the semantics $\ff{-}^\CC$ to get more abstract, official 
interpretation $\ff{-}$, which uses category $\cD$ instead of $\CC$.

\subsubsection{Semantics $\cff{-}$ in $\Fam(\CC) \rightarrow \Set$}\label{sec:semantics-C}

The interpretation is explicit about the set of variables under
which we consider types, type assignments, and terms.  Write $\Delta
\vdash \Gamma$ to mean that $\Delta\vdash\Gamma(x):\type$, for all $x$
in the domain of $\Gamma$.

The semantics of $\Delta \vdash \theta (: \type)$ and $\Delta \vdash \Gamma$
is given by a family of objects in $\CC$ indexed by the environments 
in $\ff{\Delta}$. The precise definition of $\cff{\theta}$ and $\cff{\Gamma}$ 
is given as follows: for $\eta$ in $\ff{\Delta}$,
$$
\begin{array}{rcl}
\cff{\Delta \vdash \trity{P}{Q}}_\eta 
& = & 
\tri(\ff{\Delta \vdash P}_\eta,\ff{\Delta \vdash Q}_\eta),
\\[1ex]
\cff{\Delta \vdash \theta\otimes P}_\eta 
& = & 
\inv(\cff{\Delta \vdash \theta}_\eta, \ff{\Delta \vdash P}_\eta),
\\[1ex]
\cff{\Delta \vdash \theta\rightarrow\theta'}_\eta
& = &
\cff{\Delta \vdash \theta}_\eta \Rightarrow \cff{\Delta \vdash \theta'}_\eta,
\\[1ex]
\cff{\Delta \vdash \Pi_i \theta}_\eta
& = &
\Pi_{n \in \sval} \cff{\Delta\cup \{i\} \vdash \theta}_{\eta[i\bind n]},
\\[1ex]
\cff{\Delta \vdash \Gamma}_\eta
& = &
\Pi_{x \in \dom(\Gamma)} \cff{\Delta \vdash \Gamma(x)}_\eta.
\end{array}
$$
Note that $\tri$ is used to interpret the triple type
$\trity{P}{Q}$, and $\inv$ to interpret the invariant extension
$\theta\otimes P$.

Each subtype relation $\theta \preceq_\Delta \theta'$ is
interpreted as a family of morphisms in $\CC$ of the shape
$$
  \{ \lambda x.\, x : 
    \cff{\Delta \vdash \theta}_\eta \rightarrow 
    \cff{\Delta \vdash \theta'}_\eta \}_{\eta \in \ff{\Delta}}.
$$ 
Note that every morphism in the family is implemented (or realized)
by the identity function. In order for this definition to typecheck,
the underlying cpo of the source object $\cff{\theta}_\eta$
should be included in that of the target $\cff{\theta'}_\eta$,
and the parameterized per of the source should imply that of the 
target for all instantiations. In the lemma below, we prove that both
of these requirements hold.
\begin{lem}\label{lemma:sem-subtype-C}
If a subtype relation $\theta \preceq_\Delta \theta'$ is derivable,
then for all $\eta$ in $\ff{\Delta}$, 
\begin{enumerate}
\item objects $\cff{\Delta \vdash \theta}_\eta$ and $\cff{\Delta \vdash \theta'}_\eta$ 
have the same underlying cpo, and 
\item their per parts $R$ and $R'$ satisfy that $ \forall p.\; R(p) \subseteq R'(p)$.
\end{enumerate}
\end{lem}
\proof
The proof proceeds by the induction on the derivation of 
$\theta \preceq_\Delta \theta'$. First, we consider the base cases where 
$\theta \preceq_\Delta \theta'$ is proved by an axiom.
In all the base cases except the generalized frame rule, 
objects $\cff{\theta}_\eta$ and $\cff{\theta'}_\eta$ are identical, 
because $\inv$ preserve all categorical structure used to interpret
types (Lemmas~\ref{lemma:inv-preservation}, 
\ref{lemma:inv-preservation2} and \ref{lemma:inv-preservation3}).
When $\theta \preceq_\Delta \theta'$ is derived by the generalized
frame rule, so that $\theta' = \theta \otimes P$ for some $P$,
object $\cff{\theta'}_\eta$ is $\inv(\cff{\theta}_\eta, \ff{P}_\eta)$.
Thus, by the definition of $\inv$, there exist $A$ and $R$ such that
$$
       \cff{\theta'}_\eta = (A,R(\ff{P}_\eta * -)) 
       \;\;\mbox{and}\;\; 
       \cff{\theta}_\eta = (A,R).
$$
The above two equations show that $\cff{\theta'}_\eta$
and $\cff{\theta}_\eta$ have the same underlying cpo. They also
imply the requirement for the parameterized pers, because $R(p) \subseteq
R(p*\ff{P}_\eta) = R(\ff{P}_\eta*p)$ for all $p$.

Second, we consider the case that Consequence is applied
in the last step of the derivation. In this case,
the derivation of $\theta \preceq_\Delta \theta'$
has the following shape:
$$
\prooftree
    \forall {\eta'} \in \ff{\Delta}.\;\;
    \ff{P'}_{\eta'} \subseteq \ff{P}_{\eta'} 
    \wwedge
    \ff{Q}_{\eta'} \subseteq \ff{Q'}_{\eta'} 
\justifies
    \trity{P}{Q} \preceq_\Delta \trity{P'}{Q'}
\endprooftree
$$
By the definition of the semantics of types, both $\cff{\trity{P}{Q}}_\eta$ 
and $\cff{\trity{P'}{Q'}}_\eta$ have $\comm$ as their underlying 
cpo. We will now show that their parameterised pers also satisfy the
requirement in the lemma. Let $R,R'$ be parameterized pers of
$\cff{\trity{P}{Q}}_\eta$ and
$\cff{\trity{P'}{Q'}}_\eta$, respectively.
Then, for all $p$ and $c_0,c_1 \in \comm$,
$$
\begin{array}{@{}r@{\,}c@{\,}l@{}}
  c_0[R(p)]c_1 
&
\iff
&
  (\forall h \in \ff{P}_\eta {*} p {*} \strue.\;
    c_0(h) {=} c_1(h))
  \wedge
  (c_0,c_1 \in |R(p)|)
\\[1ex]
&
\iff
&
  (\forall h \in \ff{P}_\eta {*} p {*} \strue.\;
    c_0(h) {=} c_1(h))
  \wedge
  (\forall h \in \ff{P}_\eta {*} p.\,
    (c_0(h),c_1(h) \subseteq \ff{Q} {*} p))
\\[1ex]
&
\implies
&
  (\forall h \in \ff{P'}_\eta {*} p {*} \strue.\;
    c_0(h) {=} c_1(h))
  \wedge
  (\forall h \in \ff{P'}_\eta {*} p.\,
    (c_0(h),c_1(h) \subseteq \ff{Q'} {*} p))
\\[1ex]
&
\iff
&
  (\forall h \in \ff{P'}_\eta {*} p {*} \strue.\;
    c_0(h) {=} c_1(h))
  \wedge
  (c_0,c_1 \in |R'(p)|)
\\[1ex]
&
\iff
&
  c_0[R'(p)]c_1.
\end{array}
$$
The implication above uses the assumption that
$P'$ is the strengthening of $P$ and $Q'$ is the
weakening of $Q$, and all the equivalences are simply the rolling
or unrolling of some definition. We have just shown
that $R(p) \subseteq R'(p)$ for all $p$, as required.

Third, we consider the cases of inference rules for
the type constructors, $\rightarrow$, $\Pi$
and $\otimes$.  All these cases follow from the induction hypothesis
and the definition of appropriate functors, which 
are used to interpret $\rightarrow$, $\Pi$ and $\otimes$. 
We illustrate this general pattern by proving the case of
$\rightarrow$.
Suppose that the last step of the derivation of $\theta \preceq_\Delta \theta'$ has
the form:
$$
\prooftree
    \theta'_0 \preceq_\Delta \theta_0
    \quad
    \theta_1 \preceq_\Delta \theta'_1 
\justifies
    \theta_0 \rightarrow \theta_1 
    \preceq_\Delta
    \theta'_0 \rightarrow \theta'_1 
\endprooftree
$$
For $i = 0,1$, let $(A_i,R_i) = \cff{\theta_i}_\eta$ and 
$(A'_i,R'_i) = \cff{\theta'_i}_\eta$. Then, by the induction
hypothesis, we have that
$$
    A'_0 = A_0,\;\;
    A'_1 = A_1,\;\;
    (\forall p.\, R'_0(p) \subseteq R_0(p)),\;\;\mbox{and}\;\;
    (\forall p.\, R_1(p) \subseteq R'_1(p)).
$$
So, the underlying cpos of $\cff{\theta_0 \rightarrow \theta_1}_\eta$
and $\cff{\theta'_0 \rightarrow \theta'_1}_\eta$ are the same cpo 
of continuous functions from $A_0$ to $A_1$.
The remaining requirement is to show
that $(R_0\Rightarrow R_1)(p) \subseteq (R'_0\Rightarrow R'_1)(p)$ for all $p$,
and it is proved below:
$$
\begin{array}{@{}r@{\,}c@{\,}l@{}}
   f[(R_0\Rightarrow R_1)(p)]g 
&
\iff
&
   \forall p_0.\, f[R_0(p{*}p_0) \rightarrow R_1(p{*}p_0)]g 
\;\;\hfill
(\because \mbox{Def. of $R_0\Rightarrow R_1$})
\\[1ex]
&
\implies
&
   \forall p_0.\, f[R'_0(p{*}p_0) \rightarrow R'_1(p{*}p_0)]g 
\;\;\hfill
(\because \forall q. R'_0(q) {\subseteq} R_0(q) \wedge R_1(q) {\subseteq} R'_1(q))
\\[1ex]
&
\iff
&
   f[(R'_0\Rightarrow R'_1)(p)]g
\;\;\hfill
(\because \mbox{Def. of $R'_0\Rightarrow R'_1$}).
\end{array}
$$ 

Finally, we consider the inference rule for transitivity. Suppose that the
last step of the derivation of $\theta \preceq_\Delta \theta'$ 
has the form:
$$
\prooftree
   \theta \preceq_\Delta \theta_0
   \quad
   \theta_0 \preceq_\Delta \theta'
\justifies
   \theta \preceq_\Delta \theta'
\endprooftree
$$
By the induction hypothesis, all of $\cff{\theta}_\eta$, 
$\cff{\theta_0}_\eta$ and $\cff{\theta'}_\eta$ have the same 
underlying cpos. Let $R,R_0,R'$ be parameterized
pers of $\cff{\theta}_\eta$, $\cff{\theta_0}_eta$ and $\cff{\theta'}_\eta$,
respectively. By the induction hypothesis again, we have that
$$
    \forall p.\; R(p) \,\subseteq\, R_0(p) \,\subseteq\, R'(p).
$$
We have just shown that the lemma holds in this case.\qed

Finally, we define the semantics of each typing judgment 
$\Gamma \vdash_\Delta M : \theta$ by an indexed family of morphisms
in $\CC$ of the form:
$$
  \{f_\eta : 
    \cff{\Delta \vdash \Gamma}_\eta \rightarrow \cff{\Delta \vdash \theta}_\eta \}_{\eta \in \ff{\Delta}}.
$$

\break\noindent The semantics is given by induction on the derivation of the judgment,
and it is shown in Figure~\ref{fig:semantics-terms}. The interpretation
uses the categorical structure of $\CC$ in a standard
way. The only specific parts are the interpretation of basic imperative
operations, where we use six basic semantic constants
$$
    \sskip,\;
    \sseq,\; 
    \snew,\; 
    \sread,\; 
    \sfree,\; 
    \mbox{and}\; 
    \swrite,\; 
$$ 
which are also defined in the figure.
\begin{figure*}
\begin{frameit}
$$
\begin{array}{@{}r@{\,}c@{\,}l@{}}
\cff{\Gamma,x\colon \theta\vdash_\Delta x:\theta}_\eta \rho
& = &
\rho(x)
\\[0.5ex]
\cff{\Gamma\vdash_\Delta \lambda x\colon\theta.M:\theta\rightarrow \theta'}_\eta \rho
& = &
\lambda c.\;
 \cff{\Gamma,x\colon \theta \vdash_\Delta M:\theta'}_\eta (\rho[x\bind c])
\\[0.5ex]
\cff{\Gamma\vdash_\Delta M M':\theta}_\eta \rho
& = &
  (\cff{\Gamma \vdash_\Delta M:\theta'\rightarrow \theta}_\eta\rho)\;
  (\cff{\Gamma \vdash_\Delta M':\theta'}_\eta\rho)
\\[0.5ex]
\cff{\Gamma\vdash_\Delta \lambda i.M : \Pi_i\theta}_\eta\rho
& = &
\lambda n.\;
\cff{\Gamma\vdash_{\Delta\cup\{i\}} M : \theta}_{\eta[i\bind n]} \rho 
\\[0.5ex]
\cff{\Gamma\vdash_\Delta M E : \theta[E/i]}_\eta\rho
& = &
  (\cff{\Gamma\vdash_\Delta M : \Pi_i\theta}_\eta\rho)\;
  (\ff{E}_\eta)
\\[0.5ex]
\cff{\Gamma\vdash_\Delta M:\theta'}_\eta\rho
& = &
(\cff{\theta \preceq_\Delta \theta'}_\eta)\;
(\cff{\Gamma\vdash_\Delta M:\theta}_\eta\rho)
\\[0.5ex]
\cff{\Gamma \vdash_\Delta \fix\, M : \theta}_\eta\rho
& = &
\lfix\;
(\cff{\Gamma \vdash_\Delta M : \theta\rightarrow \theta}_\eta\rho)
\\[0.5ex]
\cff{\Gamma \vdash_\Delta \ifz\,E\,M\,M' : \trity{P}{Q}}_\eta\rho
& = &
\Ifthenelse
  {(\ff{\Delta \vdash E}_\eta \,{=}\, 0)}
  {\cff{\Gamma \vdash_\Delta M : \trity{P\wedge E{=}0}{Q}}_\eta\rho}
  {\cff{\Gamma \vdash_\Delta M' : \trity{P\wedge E{\not=}0}{Q}}_\eta\rho}
\\[0.5ex]
\cff{\Gamma \vdash_\Delta M;M' : \trity{P}{Q}}_\eta\rho
& = &
\sseq\;
(\begin{array}[t]{@{}l@{}} 
  \cff{\Gamma \,{\vdash_\Delta} M {:} \trity{P}{P'}}_\eta\rho,\\
  \cff{\Gamma \,{\vdash_\Delta} M' {:} \trity{P'}{Q}}_\eta\rho)
 \end{array}
\\[0.5ex]
\cff{\Gamma \vdash_\Delta \myskip : \trity{P}{P}}_\eta\rho
& = &
\sskip\; (\bot)
\\[0.5ex]
\cff{\Gamma \vdash_\Delta \mylet{i{=}\new}{M}:\trity{P}{Q}}_\eta\rho
& = &
\snew\,
(\lambda n.\,
\cff{\Gamma \,{\vdash_{\Delta\cup\{i\}}} M {:} \trity{i{\mapsto}{-}{*}P}{Q}}_{\eta[i\bind n]}\rho
)
\\[0.5ex]
\cff{\Gamma \,{\vdash_\Delta} \mylet{i{=}[E]}{M} {:} \trity{\exists i.E{\mapsto}i{*}P}{Q}}_\eta\rho
& = &
\sread\,
\begin{array}[t]{@{}l@{}}
(\ff{\Delta \vdash E}_\eta) \\
(\lambda n.\, \cff{\Gamma \,{\vdash_{\Delta\cup\{i\}}} M {:} \trity{E{\mapsto} i{*}P}{Q}}_{\eta[i\bind n]}\rho)
\end{array}
\\[0.5ex]
\cff{\Gamma \vdash_\Delta \free(E) : \trity{E{\mapsto}{-}}{\emp}}_\eta\rho
& = &
\sfree\;(\ff{\Delta \vdash E}_\eta)\;(\bot)
\\[0.5ex]
\cff{\Gamma \vdash_\Delta [E]{:=}E' : \trity{E{\mapsto}{-}}{E{\mapsto}E'}}_\eta\rho
& = &
\swrite\;(\ff{\Delta \vdash E}_\eta, \ff{\Delta \vdash E'}_\eta)\;(\bot)
\\[0.5ex]
\end{array}
$$
where 
$\sskip$,
$\sseq$,
$\snew$,
$\sread(m)$,
$\sfree(m)$,
and
$\swrite(m,m')$
are the following morphisms in $\CC$:
$$
\begin{array}{c}
\begin{array}{rclrcl}
m{\mapsto}{-} \in \spred & \defeq & \{[m\bind n] \mid n \in \sval\}
\quad\qquad
m{\mapsto}n \in \spred & \defeq & \{[m\bind n]\}
\end{array}
\\
\\
\begin{array}{rcl}
\sskip_p  & : & 1 \rightarrow \tri(p,p)
\\
\sskip_p  & \defeq & \lambda x.\,\lambda h.\{h\}
\\[1.5ex]
\sseq_{p,p',q}  & : & \tri(p,p') \times \tri(p',q) \rightarrow \tri(p,q)
\\
\sseq         &  \defeq  & {\lambda (c,c').\, \lambda h.\,
                           \{ \wrong \mid \wrong \in c(h) \}
                           \cup
                           \bigcup \{ c'(h') \mid h' \in c(h) \}}
\\[1.5ex]
\snew_{p,q} & : & (\Pi_{n \in \sval} \tri(n{\mapsto}{-}*p,q))
                       \rightarrow
                       \tri(p,q) \\
\snew & \defeq      & {\lambda c.\,\lambda h.\, {} \bigcup\{c(n)([n\bind n']\cdot h) \mid n,n' \in \sval \;\wedge\;  n \not\in \dom(h)\}}
\\[1.5ex]
\sread(m)_{(\{p_n\}_n,q)} & : & (\Pi_{n\in \sval} \tri(m{\mapsto}n*p_n,q))
                           \rightarrow
                         \tri(\bigcup \{m{\mapsto}n*p_n\mid n \in \sval\},\;
                               q) \\
\sread(m) & \defeq      & {\lambda c.\,\lambda h.\, \ifthenelse
                              {m \in \dom(h)}
                              {c(h(m))(h)}
                              {\{\wrong\}}}
\\[1.5ex]
\sfree(m) & : & 1 \rightarrow \tri(m {\mapsto}{-}, \semp) \\
\sfree(m) & \defeq & {\lambda x.\, \lambda h.\,
                    \ifthenelse
                       {m \in \dom(h)}
                       {\{h[m\bind \undefined]\}}
                       {\{\wrong\}}}
\\[1.5ex]
\swrite(m,m') & : & 1 \rightarrow \tri(m {\mapsto} {-},
                                             m {\mapsto} m') \\
\swrite(m,m') & \defeq & \lambda x.\, \lambda h.\,
                             \ifthenelse
                             {m \in \dom(h)}
                             {\{h[m\bind m']\}}
                             {\{\wrong\}}
\end{array}
\end{array}
$$
\caption{Interpretation of Terms}
\label{fig:semantics-terms}
\end{frameit}
\end{figure*}

For this interpretation of terms, the question of well-definedness
arises, because of the introduction and elimination of dependent
function type $\Pi_i \theta$.  The semantic definition of $\lambda i.
M$ assumes that if $\Gamma$ does not contain the variable $i$, it is
interpreted as the same object in $\CC$ no matter how we change or
even drop the value of $i$ in the index. The definition of $\cff{M E}$
assumes that the reindexing precisely models the substitution. The
following lemmas show that these two assumptions indeed hold.
\begin{lem}\label{lemma:variables-type}
If \/ $i\not\in \Delta$ and $\Delta \vdash \theta$,\, 
then 
$$
 \forall \eta \in \ff{\Delta}.\, \forall n \in \sval.\; 
 \cff{\Delta \vdash \theta}_\eta = \cff{\Delta \cup \{i\} \vdash \theta}_{\eta[i\bind n]}.
$$
\end{lem}
\proof
The lemma can be proved by straightforward induction on 
the structure of $\theta$. We omit the details.
\qed
\begin{lem}\label{lemma:variables-context}
If \/ $i\not\in \Delta$ and $\Delta \vdash P$,
then 
$$
 \forall \eta \in \ff{\Delta}.\, \forall n \in \sval.\; 
 \cff{\Delta \vdash \Gamma}_\eta = \cff{\Delta \cup \{i\} \vdash \Gamma}_{\eta[i\bind n]}.
$$
\end{lem}
\proof
The lemma follows from Lemma~\ref{lemma:variables-type}, as shown below:
$$
\begin{array}{rcl}
  \cff{\Delta \cup \{i\} \vdash \Gamma}_{\eta[i\mapsto n]} 
  & 
  =
  &
  \Pi_{x \in \dom(\Gamma)}\cff{\Delta \cup \{i\} 
                              \vdash \Gamma(x)}_{\eta[i\mapsto n]}
\\
  & 
  =
  &
  \Pi_{x \in \dom(\Gamma)}\cff{\Delta \vdash \Gamma(x)}_\eta
\qquad
  (\because~  \mbox{Lemma~\ref{lemma:variables-type}})
\\
  &
  = 
  &
  \cff{\Delta \vdash \Gamma}_\eta.
\end{array}
$$
\qed
\begin{lem}
If \/ $i\not\in \Delta$,\, $\Delta \cup \{i\} \vdash \theta$,\,
and $\Delta \vdash E$,\, 
then 
$$
   \forall \eta \in \ff{\Delta}.\; 
   \cff{\Delta \vdash \theta[E/i]}_\eta = \cff{\Delta \cup\{ i \}\vdash \theta}_{\eta[i\bind \ff{E}_\eta]}.
$$
\end{lem}
\proof
This lemma holds because the reindexing of the family fibration 
$\Fam(\CC) \rightarrow \Set$ preserves on the nose all the categorical structure 
that is used to interpret types. A more concrete, direct proof can be obtained by 
induction on the structure of $\theta$. We omit the details.
\qed

\subsubsection{Semantics $\ff{-}$ in $\Fam(\cD) \rightarrow \Set$}
The official semantics $\ff{-}$ of the language uses the fibration 
$\Fam(\cD) \rightarrow \Set$, rather than $\Fam(\CC) \rightarrow \Set$. 
It is obtained by applying the embedding functor $E \colon \CC\rightarrow \cD$ 
to the semantics $\cff{-}$ of the previous section. Concretely, the 
semantics $\ff{-}$ is defined as follows: for all $\eta \in \ff{\Delta}$,
$$
\begin{array}{rcrcl}
  \ff{\Delta \vdash \theta}_\eta 
& = & 
  E(\cff{\Delta \vdash \theta}_\eta)
& = & 
  \cff{\Delta \vdash \theta}_\eta,
\\[1ex]
  \ff{\Delta \vdash \Gamma}_\eta 
& = & 
  E(\cff{\Delta \vdash \Gamma}_\eta)
& = & 
  \cff{\Delta \vdash \Gamma}_\eta,
\\[1ex]
  \ff{\theta \preceq_\Delta \theta'}_\eta 
& = & 
\multicolumn{3}{l}{
  E(\cff{\theta \preceq_\Delta \theta'}_\eta),
}
\\[1ex]
  \ff{\Gamma \vdash_\Delta M : \theta}_\eta 
& = & 
\multicolumn{3}{l}{
  E(\ff{\Gamma \vdash_\Delta M : \theta}_\eta). 
}
\end{array}
$$
Note that in the first two equations, we use the fact that $E$
is the identity on objects.

We point out that $\ff{-}$ can be presented in a compositional
style, using the categorical structure 
of the fibration $\Fam(\cD)\rightarrow \Set$.\footnote{The conference
version of this paper defined $\ff{-}$ in such a  style.}
In that presentation,
the types are interpreted 
using exponentials, small products, $\inv'$ and $\tri'$ for $\cD$;
and the terms are interpreted by appropriate categorical 
combinators and the embedding of the six constants in 
Figure~\ref{fig:semantics-terms}. This direct definition of $\ff{-}$
is identical to the semantics in this section, because
the embedding functor $E$ preserves all the relevant categorical 
structure (Lemma~\ref{lemma:preservation-E}).

\subsection{Adequacy}
\label{sec:adequacy}
Our semantics of terms needs further justification in two ways. First,
the interpretation of a typing judgment needs to be shown coherent.
The interpretation is defined over a proof derivation of the judgment,
so two different derivations of the same judgment might have different
denotations. This is troublesome for us especially, because our goal
is to give a semantics of a programming language with a
separation-logic type system, instead of a semantics of a proof in
separation logic. Second, the connection with the standard semantics
needs to be provided. Our semantics uses subsumption which never
arises in the standard interpretation. Thus, it could
be substantially different from the standard interpretation.  In this
section, we provide justification for both of these two issues.

We consider another interpretation $\sff{-}$ of our language, called
standard interpretation, which ignores all
assertions in the types. In the standard interpretation,
$\trity{P}{Q}$ means the same thing no matter what $P$ and $Q$ are,
and for all $P$, $\theta\otimes P$ and $\theta$ have identical
interpretations. Let $\tri''$ be the constant functor from
$\cP^\op\times \cP$ to $\CPO$ such that $\tri''(p,q) = \comm$, and let
$\inv''$ be a functor given by the first projection from $\CPO\times
\cPr$ to $\CPO$. The standard interpretation is the interpretation in 
Section~\ref{sec:semantics-C}, where we use $\CPO$, $\tri''$
and $\inv''$ instead of $\CC$, $\tri$ and $\inv$.  It interprets
types and type assignments just like the interpretation in
Section~\ref{sec:semantics-C}, but it uses functors on
$\CPO$, instead of those on $\CC$. 
\begin{lem}\label{lemma:coherence-standard}
If a subtype relation $\theta \preceq_\Delta \theta'$ is derivable, then
$\theta$ and $\theta'$ have the identical denotation in the standard 
interpretation.
\end{lem}
\proof
We prove the lemma by induction on the derivation of 
the subtype relation $\theta \preceq_\Delta \theta'$. First,
we consider the case that the subtype relation is derived by an axiom.
In all six axioms, $\theta$ and $\theta'$ are both Hoare-triple types,
or they are different only for the invariant added by $\otimes$.
Note that in the standard interpretation, all triple types mean the
same cpo $\comm$ and the added invariants by $\otimes$ are ignored.
Thus, we have that $\sff{\theta}_\eta = \sff{\theta'}_\eta$ for all 
environments $\eta \in \ff{\Delta}$. Next, we consider the cases where some inference rule
 is applied at the last step of the derivation. Pick an environment $\eta$
in $\ff{\Delta}$. If the last rule in the derivation 
is Consequence, both $\theta$ and $\theta'$ are Hoare-triple objects,
so $\sff{\theta}_\eta$ and $\sff{\theta'}_\eta$ are the same cpo $\comm$.
If the last applied rule is 
an inference rule other than Consequence, 
$\sff{\theta}_\eta$ and $\sff{\theta'}_\eta$ 
are obtained by applying the same functor on the denotations 
of their subparts. By applying the induction hypothesis to these
subparts, we can prove the lemma. For instance, if the last applied rule 
is the structural rule for $\rightarrow$, there are $\theta_0,\theta'_0,\theta_1,\theta'_1$
such that
$$
      \theta = \theta_0\rightarrow \theta_1,\;\;
      \theta = \theta'_0\rightarrow \theta'_1,\;\;
      \theta'_0 \preceq_\Delta \theta_0,\;\;\mbox{and}\;\;
      \theta_1 \preceq_\Delta \theta'_1.
$$
By the induction hypothesis, $\sff{\theta_i}_\eta = \sff{\theta'_i}_\eta$ for \/ $i=0,1$.
This implies that $\sff{\theta}_\eta$ and $\sff{\theta}_\eta$ are identical.
\qed

The standard interpretation defines
the meaning of typing judgments $\Gamma \vdash_\Delta M : \theta$,
by repeating the clauses in Figure~\ref{fig:semantics-terms}.
Although the interpretation is given inductively on the typing derivation,
Lemma~\ref{lemma:coherence-standard} ensures that 
$\sff{\Gamma \vdash_\Delta M : \theta}$ does not depend 
on derivations, because it guarantees that
$\sff{\theta \preceq_\Delta \theta'}$ is the identity morphism.
As usual, we can give the operational semantics,
and prove the computational adequacy of the standard interpretation.
Since this is completely standard, we omit it.
 
The standard interpretation is closely related to the semantics $\cff{-}$
in Section~\ref{sec:semantics-C}. Note that from
the category $\CC$ to $\CPO$, there is a forgetful functor $F$ that
maps an object $(A,R)$ to $A$, and a morphism $f$ to $f$.  This
forgetful functor preserves all the categorical structure of $\CC$
that we use to interpret the types of our language:
\begin{lem}\label{lemma:preservation-F}
$F$ is a faithful functor that preserves the CCC structure and
the small products of $\CC$, and makes the following diagrams commute.
$$
\xymatrix{
\CC \times \cPr   \ar[r]^\inv  \ar[d]_{F\times\Id} & \CC \ar[d]^F \\
\CPO \times \cPr   \ar[r]^-{\inv''} & \CPO
}
\qquad
\xymatrix{
\cP^\op \times \cP  \ar[r]^\tri \ar[d]_{\Id} & \CC \ar[d]^F \\
\cP^\op \times \cP  \ar[r]^-{\tri''} & \CPO
}
$$
\end{lem}
\proof
First, we prove that the forgetful functor $F$ preserves the exponentials 
and small products of $\CC$. For this, it is sufficient
to prove the preservation of
four elements:
exponential objects, small product objects, evaluation morphisms,
and projection morphisms.
Note that both the CCC structure and small products of $\CC$ are defined
using the corresponding structure of $\CPO$; the 
first components of exponential objects and small product objects
of $\CC$ are defined by  exponential objects and small product objects 
of $\CPO$, and evaluation morphisms and projection morphisms in $\CC$ are
precisely evaluation morphisms and projection morphisms in $\CPO$. 
Since $F$ projects the first component of each object in $\CC$
and maps each morphism in $\CC$ to itself, it
preserves the required four elements.
For instance, for all objects $(A,R),(B,S)$ 
in $\CC$, the first component of their exponential $(A,R) \Rightarrow (B,S)$ 
is the cpo $A \Rightarrow B$ of continuous functions from $A$
to $B$, which is precisely the exponential of $A$ and $B$ in $\CPO$.
Thus, $F((A,R)\Rightarrow (B,S))$ is $F(A)\Rightarrow F(B)$.

Next, we prove that the diagram for $\inv$ and $\inv''$ commutes.
Since $\inv''$ is the projection of the first component, 
$\inv'' \circ (F\times \Id)$ is $F \circ \myfst$. So, it suffices
to show that $F \circ \myfst = F \circ \inv$. Consider objects 
$((A,R),p)$, $((B,S),q)$ and
a morphism $(f, p \sqsubseteq q)\colon ((A,R),p) \rightarrow ((B,S),q)$
in $\CC \times \cPr$. Then, 
$$
\begin{array}{l}
  (F \circ \inv)((A,R),p) \;=\; F (A,R(p*-)) 
                          \;=\; A \;=\; (F \circ \myfst)((A,R),p),
  \;\;\mbox{ and }\;\;
\\
  (F \circ \inv)(f, p \sqsubseteq q)
  \;=\; F(f) \;=\; f
  \;=\; (F \circ \myfst)(f, p \sqsubseteq q).
\end{array}
$$
Thus, $F \circ \myfst = F \circ \inv$, as required.

Finally, we prove the commutative diagram for $\tri$ and $\tri''$.
Consider objects (or predicate pairs) $(p,q),(p',q')$ and a
morphism $(p'\subseteq p,q\subseteq q')\colon (p,q) \rightarrow (p',q')$
in $\cP^\op \times \cP$. Then,
$$
   (F \circ \tri)(p,q) \;=\; F([p,q]) \;=\; \comm
   \;\;\mbox{ and }\;\;
   (F \circ \tri)(p'\subseteq p,q\subseteq q') \;=\; \id.
$$
Thus, $F \circ \tri$ is the constant functor to $\comm$, 
so it is identical to $\tri''$.
\qed
Lemma~\ref{lemma:preservation-F} implies that the interpretation
of types in $\CPO$ factors through the interpretation in
$\CC$. The following lemma show that the interpretation of 
terms has a similar property.
\begin{prop}\label{prop:preservation-semantics}
  The functor $F\colon \CC \rightarrow \CPO$ preserves the
  interpretation of terms: for all typing judgments $\Gamma
  \vdash_\Delta M : \theta$ and all $\eta \in \ff{\Delta}$,
$$
F(\cff{\Gamma \vdash_\Delta M : \theta}_\eta) = 
\sff{\Gamma \vdash_\Delta M : \theta}_\eta.
$$
\end{prop}
\proof
Pick an arbitrary $\eta \in \ff{\Delta}$ and choose any
$\rho' \in \sff{\Gamma}_\eta$. Then,
$$
F(\cff{\Gamma \vdash_\Delta M : \theta}_\eta)(\rho') = 
\cff{\Gamma \vdash_\Delta M : \theta}_\eta\;\rho',
$$ 
because $F(f)$ only changes the ``type'' of $f$, not
the implementation of $f$. Thus, it is sufficient to show that
$$
\forall \eta,\rho'.\;\,
  \cff{\Gamma \vdash_\Delta M : \theta}_\eta\;\rho' =
  \sff{\Gamma \vdash_\Delta M : \theta}_\eta\;\rho'.
$$
We prove this equality by induction on the derivation of 
$\Gamma \vdash_\Delta M : \theta$. Since $\sff{-}$ and $\ff{-}$ use 
the same clauses to define the meaning of 
$\Gamma \vdash_\Delta M \colon \theta$,  the induction easily goes through in 
all cases.  For instance, consider the case 
where the subsumption rule is applied
at the last step of the derivation. For all 
environments $\eta \in \ff{\Delta}$ and all $\rho' \in \sff{\Gamma}_\eta$,
$$
\begin{array}{@{}r@{\;}c@{\;}l@{}}
\cff{\Gamma \vdash_\Delta M : \theta}_\eta\;\rho'
& = &
\cff{\theta_0 \preceq_\Delta \theta}_\eta\;
(\cff{\Gamma \vdash_\Delta M : \theta_0}_\eta\;\rho')
\\[0.5ex]
& = &
\cff{\Gamma \vdash_\Delta M : \theta_0}_\eta\;\rho'
\qquad\hfill
(\because~ \cff{\theta_0\preceq_\Delta \theta}_\eta\, x = x)
\\[0.5ex]
& = &
 \sff{\Gamma \vdash_\Delta M : \theta_0}_\eta\;\rho'
\qquad\hfill
(\because~ \mbox{Induction Hypothesis})
\\[0.5ex]
& = &
   (\sff{\theta_0 \preceq_\Delta \theta}_\eta) \circ
   (\sff{\Gamma \vdash_\Delta M : \theta_0}_\eta\;\rho')
\qquad\hfill
(\because~ \sff{\theta_0 \preceq_\Delta \theta}_\eta\;x = x)
\\[0.5ex]
& = &
 \sff{\Gamma \vdash_\Delta M : \theta}_\eta.
\end{array}
$$
\qed

Recall that the official semantics $\ff{-}$ of our language is obtained by
applying the full functor $E$ to the semantics $\cff{-}$,
and that the functor $F$ is faithful. Together with these facts,
Lemma~\ref{lemma:preservation-F} and Proposition~\ref{prop:preservation-semantics} 
show that the official semantics $\ff{-}$ is obtained from the standard interpretation
$\sff{-}$ by first selecting some elements, and then quotienting those selected elements. 
\begin{cor}
  The semantics $\ff{-}$ is coherent: the semantics of a typing judgment does
  not depend on derivations.
\end{cor}
\proof
  Let $\cP_1$ and $\cP_2$ be two derivations of a judgment $\Gamma
  \vdash_\Delta M : \theta$. We note that the standard semantics is
  coherent; only the subsumption rule is not syntax-directed, but in
  the standard semantics, this rule does not contribute to the
  interpretation, because all the subtype relations 
  $\theta \preceq_\Delta \theta'$ denote the family of identity
  morphisms. Thus, for all environments $\eta \in \ff{\Delta}$, we have
$$
   \ff{\cP_1}^\CPO_\eta = \ff{\cP_2}^\CPO_\eta.
$$
Then, by Proposition~\ref{prop:preservation-semantics} and the faithfulness
of $F$,
$$
\begin{array}{rcl}
\ff{\cP_1}^\CPO_\eta = \ff{\cP_2}^\CPO_\eta
&
\Longrightarrow
&
F(\ff{\cP_1}^\CC_\eta)
=
F(\ff{\cP_2}^\CC_\eta)
\\
&
\Longrightarrow
&
\ff{\cP_1}^\CC_\eta
=
\ff{\cP_2}^\CC_\eta
\qquad\hfill
(\because~  \mbox{$F$ is faithful})
\\
&
\Longrightarrow
&
E(\ff{\cP_1}^\CC_\eta)
=
E(\ff{\cP_2}^\CC_\eta)
\\
&
\Longrightarrow
&
\ff{\cP_1}_\eta
=
\ff{\cP_2}_\eta
\qquad\hfill
(\because~  \mbox{Definition of $\ff{-}$}).
\end{array}
$$
\qed

\section{Conjunction Rule}
\label{sec:conjunction-rule}
The conjunction rule is often omitted from Hoare logic,
but it is a useful proof rule that lets one combine two 
Hoare triples about a single command.
In our type system, it can be expressed as follows:
$$
\prooftree
  \Gamma \vdash_\Delta M : \trity{P}{Q} 
  \quad
  \Gamma \vdash_\Delta M : \trity{P'}{Q'} 
\justifies
  \Gamma \vdash_\Delta M : \trity{P \wedge P'}{Q \wedge Q'} 
\endprooftree
$$

Unfortunately, we cannot immediately include the conjunction rule in our
type system.  In \cite{yang-ohearn-reynolds-popl04}, Reynolds has proved
that if a proof system contains the conjunction rule and the second-order
frame rule, together with Consequence and the ordinary (first-order) frame
rule, then the system becomes inconsistent. More specifically, Reynolds's
result implies that once the conjunction rule is added to our type system,
we can derive 
$\vdash_{\{\}} \myskip : \trity{(\exists x,y. x{\mapsto}y)*\true}{\false}$, 
which incorrectly expresses that $\myskip$ diverges when
the input heap is not empty.

In the case of the second-order frame rule, several solutions have been
proposed to overcome this problem.  In this section we adopt one of the
proposals, modify the separation-logic type system accordingly, and extend
the modified system with the conjunction rule. Then, we define an
adequate semantics of the new type system, thereby showing that all the
higher-order frame rules can be used with the conjunction rule, as long as
the frame rules add only precise invariants.

We recall the definition of precise predicates in separation logic
\cite{yang-ohearn-reynolds-popl04}. 
A predicate $p$ is precise if and only if for every heap $h$, there
is at most one subheap $h_0$ of $h$ (i.e., $h_0\cdot h_1 = h$ for
some $h_1$) such that $h_0 \in p$. We also call
an assertion $\Delta \vdash P$ {\em precise\/}
 when $\ff{P}_\eta$ is a precise predicate
for all $\eta \in \ff{\Delta}$.

The proposal that we use is to restrict the second-order frame rule such
that it is used with only precise assertions.  We adopt the proposal in our
separation-logic type system by limiting the second parameter of the type
constructor $\otimes$ to precise assertions. Note that in the resulting
restricted type system, only precise assertions can be added as invariants,
because the generalized frame rule $\theta \preceq_\Delta \theta\otimes P$
is applicable only with a precise assertion $P$. Thus, the second or third
order frame rule can add only precise assertions as invariants.  We may
then extend the restricted type system with the conjunction rule.  Note
that the result of this extension, denoted $\T$, includes the conjunction
rule and all (restricted) higher-order frame rules.  In the remainder of
this section, we focus on giving an adequate semantics of $\T$.

Before giving the semantics of $\T$, we point out
that requiring invariants to be precise is
not as restrictive as it seems; all the examples
in Section~\ref{sec:example-proofs} use precise invariants
only, so they typecheck in $\T$.

The semantics of the type system $\T$ is given by categories
$\CC_0$ and $\cD_0$. The category $\CC_0$ is identical to $\CC$,
except that the per component of each object is parameterized 
by {\em precise\/} predicates, instead of all predicates. An object in $\CC_0$
is a pair of cpo $A$ and parameterized per $R$, such that 
(1) the parameterization of $R$ is over {\em precise\/} predicates, and
(2) for all {\em precise\/} predicates $p,q$, the per $R(p)$ 
implies $R(p*q)$, i.e., $R(p) \subseteq R(p*q)$. A morphism 
$f\colon (A,R) \rightarrow (B,S)$ in $\CC_0$ is a continuous function
>From $A$ to $B$ that maps $R(p)$-related elements to $S(p)$-related
elements for all precise $p$. The other category $\cD_0$ is constructed
 by quotienting morphisms in $\CC_0$, in the same way as $\cD$
is constructed from $\CC$.

The categories $\CC_0$ and $\cD_0$ have all the categorical structure 
that we have used in the semantics in Section~\ref{sec:semantics}.
They are cartesian closed categories with all
the small products, and they have functors for invariant extension
and Hoare triples. The only subtlety is that 
the preorder $\cPr$, which is used for functors 
for invariant extension in Section~\ref{sec:semantics}, is now replaced 
by the preorder of precise predicates with the following order 
$\sqsubseteq_p$: for all precise predicates $p,q$,
$$
   p \sqsubseteq_p q \iff
   \mbox{there exists a precise $r$ such that $p*r = q$}.
$$
This categorical structure is preserved by the functors for invariant
extension, the forgetful functor $F_0\colon \CC_0\rightarrow
\CPO$, and the quotienting functor $E_0\colon \CC_0\rightarrow \cD_0$,
in the way expressed by Lemmas~\ref{lemma:inv-preservation},
\ref{lemma:preservation-F}, and \ref{lemma:preservation-E}. 
All the definitions and results in 
Section~\ref{sec:interpretation-pl} and \ref{sec:adequacy}
can easily be transferred to $\CC_0$ and $\cD_0$,
as long as they are concerned with $\T$ without the conjunction rule. 
We now explain how to soundly interpret the conjunction rule.

Define a continuous function $\con$ from $\comm\times\comm$ 
to $\comm$ as follows:
$$
\begin{array}{rcl}
  \wrong \in \con(c,c')(h) 
&
  \iff
&
  \wrong \in c(h) \cup c'(h) 
\\
  h' \in \con(c,c')(h) 
&
  \iff
&
   h' \in c(h) \cap c'(h) 
\end{array}
$$
Function $\con$ is the key element in our interpretation of the 
conjunction rule. Intuitively, $\con(c,c')$ is a command that
is better than $c$ and $c'$: it
satisfies more Hoare triples than $c$ and $c'$, 
as long as we consider triples with sufficiently strong preconditions,
those which ensure that both $c$ and $c'$ run without generating $\wrong$. 
\begin{lem}\label{lemma:con-welldefined}
Function $\con$ is well-defined. In particular,
for all $(c,c') \in \comm\times\comm$, $\con(c,c')$ satisfies
the safety monotonicity and frame property. 
\end{lem}
\proof
The continuity follows from the fact that $\con(c,-)$ and
$\con(-,c)$ preserve arbitrary nonempty unions. Here
we focus on proving that $\con$ is a well-defined function. Pick
$(c,c') \in \comm\times\comm$. To prove that $\con(c,c') \in \comm$,
we should show that $\con(c,c')$ satisfies the safety monotonicity
and the frame property.
\begin{itemize}
\item {\em Safety Monotonicity:} Consider heaps $h_0,h_1$ such that
      $\wrong \not\in \con(c,c')(h_0)$ and $h_0\# h_1$. Then,
      $\wrong$ is neither in $c(h_0)$ nor in $c'(h_0)$. Thus,
      by the safety monotonicity of $c$ and $c'$, we have that
      $\wrong \not\in c(h_0\cdot h_1)$ and $\wrong \not\in c'(h_0\cdot h_1)$.
      This implies that $\wrong \not\in \con(c,c')(h_0\cdot h_1)$, as required.
\item {\em Frame Property:} Suppose that
      $h_0\# h_1$, $\wrong \not\in \con(c,c')(h_0)$, and
      $h' \in \con(c,c')(h_0\cdot h_1)$. Note that while
      proving the previous item, we have shown two facts: 
      (1) $\con(c,c')(h_0\cdot h_1)$ does not contain $\wrong$, 
      and (2) neither $c(h_0)$ nor $c'(h_0)$ contains
      $\wrong$. The first fact implies that
      $\con(c,c')(h_0\cdot h_1) =
      c(h_0\cdot h_1) \cap c'(h_0\cdot h_1)$, because by the definition of
      $\con$,
$$
      \begin{array}{r@{}c@{}l}
      c(h_0\cdot h_1) \cap c'(h_0\cdot h_1) 
      &
      \;\;\subseteq\;\;
      & 
      \con(c,c')(h_0\cdot h_1)  
      \\
      &
      \;\;\subseteq\;\;
      &
      \bigl(c(h_0\cdot h_1) \cap c'(h_0\cdot h_1)\bigr) \cup \{\wrong\}.
      \end{array}
$$
      Since $h'$ is in $\con(c,c')(h_0\cdot h_1)$ and 
      $\con(c,c')(h_0\cdot h_1) = c(h_0\cdot h_1) \cap c'(h_0\cdot h_1)$, 
      heap $h'$ is in $c(h_0\cdot h_1)$ as well as 
      in $c'(h_0\cdot h_1)$. Moreover, by the second fact proved
      in the previous item,
      $\wrong \not\in c(h_0)$ and $\wrong \not\in c'(h_0)$.
      Thus, we can apply the frame property of $c$ and $c'$ here.
      Once the property is applied, we obtain subheaps $h'_0,h''_0$ of $h'$
      such that
$$
      h'_0\cdot h_1 = h''_0\cdot h_1 = h'
             \;\;\wedge\;\;
              h'_0 \in c(h_0)
             \;\; \wedge\;\;
      h''_0 \in c'(h_0).
$$
     Note that the equalities force
     $h'_0$ and $h''_0$ to be the same. So, $h'_0$ should
     be in $c(h_0)\cap c'(h_0) = \con(c,c')(h_0)$. We have
     just proved that $h'_0$ is the heap required by
     the frame property of $\con(c,c')$.
\end{itemize}
\qed

For all predicates $p,q$, define an object $[p,q]$ in $\CC_0$ just
like the corresponding triple object in $\CC$, except that 
the second component of $[p,q]$ is a family of pers indexed by precise
predicates. The following lemma expresses that $\con$ properly
models a semantic version of the conjunction rule in $\CC_0$.
\begin{lem}\label{lemma:con-conjunction-rule}
For all predicates $p,q,p',q'$, 
function $\con$ is a morphism in $\CC_0$
that has type $[p,q]\times [p',q'] \rightarrow [p\cap p',q\cap q']$.
\end{lem}
\proof
Let $R,S,T$ be pers parameterized by precise predicates,
such that
$$
  (\comm,R) = [p,q],\quad
  (\comm,S) = [p',q'], \quad\mbox{ and }\quad
  (\comm,T) = [p\cap p',q \cap q'].
$$
Because of Lemma~\ref{lemma:con-welldefined}, $\con$ is a well-defined
continuous function from $\comm\times\comm$ to $\comm$. Thus, it suffices
to show that for all precise predicates $r$,
$$
       \con[R(r)\times S(r) \rightarrow T(r)]\con.
$$
Consider precise predicate $r$, and command pairs $(c_0,c'_0),(c_1,c_1')$,
such that 
$$
       (c_0,c'_0)[R(r)\times S(r)](c_1,c_1').
$$ 
First, we show that
$\con(c_0,c'_0)$ and $\con(c_1,c_1')$ are in the domain of per $T(r)$.
We focus on $\con(c_0,c'_0)$, because $\con(c_1,c_1') \in |T(r)|$
can be proved similarly.
Pick a heap $h$ in $(p\cap p')*r$. Then, $h$ is in $p*r$ and $p'*r$.
Note that $c_0$ and $c'_0$ are in $|R(r)|$ and $|S(r)|$,
and $R$ and $S$ are the per components of
$[p,q]$ and $[p',q']$. Thus, neither $c_0(h)$ nor $c'_0(h)$ contains
$\wrong$,  $c_0(h) \subseteq q*r$, and $c'_0 \subseteq q'*r$.
Thus,
$$
   \con(c_0,c'_0)(h) \;\;=\;\; c_0(h)\cap c'_0(h) 
                     \;\;\subseteq\;\; p*r \cap q'*r
                     \;\;=\;\; (p\cap q') * r.
$$
The first equality follows from the definition of $\con$, 
because $\wrong \not\in c_0(h)$ and $\wrong \not\in c'_0(h)$.
And the last equality holds, because
for all precise predicates $r_0$, 
$-*r_0$ distributes over $\cap$. 

Next, we show that $\con(c_0,c'_0)$ and $\con(c_1,c_1')$ are
$T(r)$-related. Since both $\con(c_0,c'_0)$ and $\con(c_1,c_1')$ are
in $|T(r)|$, it is enough to prove that
$$
 \forall h \in (p\cap p')*r*\strue.\;\,
     \con(c_0,c'_0)(h) = \con(c_1,c'_1)(h).
$$
Pick $h$ from $(p\cap p')*r*\strue$. Then, $h \in p*t*\strue$
and $h \in p'*t*\strue$. Since $c_0[R(p)]c_1$
and $c'_0[S(p)]c'_1$, these two membership relations of $h$
imply that none of $c_0(h),c_1(h),c_0'(h),c_1'(h)$ contains $\wrong$,
$c_0(h) = c_1(h)$, and $c'_0(h) = c'_1(h)$.
Thus,
$$
  \con(c_0,c'_0)(h) \,\;=\;\, c_0(h)\cap c'_0(h) \,\;=\,\;
                      c_1(h)\cap c'_1(h) \,\;=\;\,  \con(c_1,c'_1)(h).
$$
Since none of $c_0(h),c_1(h),c_0'(h),c_1'(h)$ contains $\wrong$,
the first and last equalities follow from the definition of $\con$.
\qed

The conjunction rule
$$
\prooftree
  \Gamma \vdash_\Delta M : \trity{P}{Q} 
  \quad
  \Gamma \vdash_\Delta M : \trity{P'}{Q'} 
\justifies
  \Gamma \vdash_\Delta M : \trity{P \wedge P'}{Q \wedge Q'} 
\endprooftree
$$
is now interpreted as follows:
$$
\ff{\Gamma \vdash_\Delta M : \trity{P \wedge P'}{Q \wedge Q'}}^X_\eta
= 
\con' \circ 
  \langle \ff{\Gamma \vdash_\Delta M : \trity{P}{Q}}^X_\eta,\; 
          \ff{\Gamma \vdash_\Delta M : \trity{P'}{Q'}}^X_\eta 
  \rangle
$$
where $X$ is $\CC_0$, $\cD_0$ or $\CPO$.
The standard semantics in $\CPO$ and the filtering semantics
in $\CC_0$ uses $\con$ for $\con'$, and in a direct-style presentation,
the quotienting semantics in $\cD_0$ uses the equivalence class $[\con]$ for
$\con'$. Note that in the standard semantics,
the conjunction rule is interpreted as the identity, because $\con \circ
\langle f,f\rangle = f$, for all morphisms $f$ in $\CPO$.

Since $E'$ and $F'$ preserve the semantic entities for $\con'$,
they preserve the interpretation of terms in the three semantics.
>From this preservation, the coherence of the quotienting semantics follows. Moreover,
since the conjunction rule means the identity in the standard semantics, the
 preservation of interpretations also implies that the conjunction rule
is always implemented by the identity function in all three semantics, 
thereby reflecting the fact that the rule does not
have any computational meaning.

\section{Related Work}
\label{sec:related-future-work}


The (first order) frame rule was discovered in the early days of
separation logic~\cite{ohearn-ishtiaq-popl01}, and it was a main
reason for the success of that logic. For example, it was vital in the
proofs of garbage collection algorithms in~\cite{yang-thesis01} and
\cite{birkedal-torpsmith-reynolds-popl04}. Recently, the second-order
frame rule, which allows reasoning about simple first-order modules,
was discovered~\cite{yang-ohearn-reynolds-popl04}. This naturally
encouraged the question of whether there are more general frame rules
that apply to higher types.

Other type systems which track state changes have been proposed in the
work on typed assembly languages
\cite{morriset-walker-crary-glew99:f_to_tal,
  ahmed-jia-walker-hierar-sto-03, tan-appel-semantics-tal-04}.
Their main focus is to obtain sound rules for proving the safety of
programs. Thus, they mostly use easy-to-define conventional
operational semantics, and prove the soundness of the proof system
syntactically (i.e., by subject reduction and progress lemmas), or
``logically'' \cite{tan-appel-semantics-tal-04}: each type is
interpreted as a subset of a single universe of ``meanings,'' and a
typing judgment is interpreted as a specification for the behavior of
programs, like a Hoare triple in separation logic.  Our
separation-logic type system is more refined in that it allows the
full power of separation logic in the types and, moreover, we also
treat higher-order procedures.

The semantics of idealized algol has been studied intensively
\cite{oles-thesis,reynolds-essence,ohearn-tennent-parametricity,yang-reddy04}.
Normally, the semantics is parameterized by the shape of the memory.
The indexing in the fibration in our semantics follows this tradition,
and it models the shape of the stack. However, the other indexing of
our semantics, the indexing by invariant predicates over heaps, has
not been used in the literature before.

The construction of the category $\cD$ is an instance of the Kripke
quotient by Mitchell and Moggi \cite{mitchell-moggi-kripkemodel91}.
The families of pers in $\cD$ form a Kripke logical relation on $\CPO$
indexed by the preorder category $\cPr$; our condition on each family
ensures that the requirement of Kripke monotonicity holds.  This
Kripke logical relation produces $\cD$ by Mitchell and Moggi's
construction.

The idea of proving coherence by relating two languages comes from
Reynolds \cite{reynolds01b}.  Reynolds proved the coherence of the
semantics of typed lambda calculus with subtyping, by connecting it
with the semantics of untyped lambda calculus.  We use the general
direction of Reynolds's proof, but the details of our proof are quite
different from Reynolds's, because we consider very different
languages.

\section{Conclusion and Future Directions}
\label{sec:conc}
We have presented a type system for idealized algol extended with
heaps that includes separation-logic specifications as types and,
moreover, defined the coherent semantics of idealized algol typed with
this system.

One shortcoming of our type system is that the higher-order frame
rules in the system allow only static modularity
\cite{parkinson-bierman-popl05}.  With the higher-order frame rules
alone, we cannot capture all the the information hiding aspect of
dynamically allocated data structures as needed for modeling abstract
data types.  However, it is well-known that abstract data types can be
modeled using existential types and we are currently considering to
enrich the assertion language with predicate variables, as in the
recently introduced higher-order version of separation
logic~\cite{biering-birkedal-torpsmith-esop05}, and to extend the
types with dependent product and sums over {\em predicates}.

Yet another future direction is to define a parametric model. Uday Reddy
pointed out that separation-logic types should validate stronger
reasoning principles for data abstraction than ordinary types, because
they let us control what clients can access more precisely.
Formalizing his intuition is the goal of the parametricity semantics.
We currently plan to use category $\CC'$ which replaces each {\em
  predicate-indexed\/} family of {\em pers\/} in $\CC$ by a {\em
  relation-indexed\/} family of {\em saturated relations}: an object
in $\CC'$ is a cpo paired with a family $T$ of {\em binary\/}
relations such that (1) $T$ is indexed by a ``typed'' relation
$r\colon p\leftrightarrow q$ on heaps (i.e., $r \subseteq p\times q$);
(2) for each predicate $p$, $T$ at the diagonal relation $\Delta_p$ is
a per; (3) for all $r\colon p\leftrightarrow q$, $T(r)$ is a saturated
relation between pers $T(\Delta_p)$ and $T(\Delta_q)$; (4)
$T(r) \subseteq T(r*r')$. The morphisms in $\CC'$ are continuous functions
that preserve the families of relations. This category has all the
categorical structure of $\CC$ that we used in the semantics of
this paper. However, it is difficult to interpret the triple types such 
that the memory allocator $\new$ lives in the category. Overcoming this problem 
will be the focus of our research in this direction.


Finally, we would like to extend the relational separation logic
\cite{yang-relational-separation-logic} to higher-order, following the
style of system $\mathcal{R}$ \cite{abadi-formal}, and we want to explore
the Curry-Howard correspondence of our type system with specification
logic \cite{reynolds-specification-logic}.

\section*{Acknowledgements}
We have benefitted greatly from discussions with Uday Reddy, Peter O'Hearn,
and David Naumann. We would like to thank anonymous referees and
Rasmus Lerchedahl Petersen for providing
useful suggestions, which in particular helped us to improve the presentation 
of the paper.  Yang was supported by grant No. R08-2003-000-10370-0
>From the Basic Research Program of the Korea Science $\&$ Engineering
Foundation. Yang, Birkedal and Torp-Smith were supported by Danish
Technical Research Council Grant 56-00-0309.


\end{document}